\documentstyle[prd,aps,epsf]{revtex}

\def\lsim{\raise0.3ex\hbox{$<$\kern-0.75em\raise-1.1ex\hbox{$\sim$}}}
\def\gsim{\raise0.3ex\hbox{$>$\kern-0.75em\raise-1.1ex\hbox{$\sim$}}}

\newcommand{\pslash}{p\kern-1ex /}
\newcommand{\Dslash}{{\cal D}\kern-1.5ex /}

\newcommand{\J}[4]{{#1} {\bf #2} (#3) #4}

\newcommand{\NP}{Nucl.~Phys.}
\newcommand{\NPSup}{Nucl.~Phys.~B (Proc.~Suppl.)}
\newcommand{\PL}{Phys.~Lett.}
\newcommand{\PR}{Phys.~Rev.}
\newcommand{\PRL}{Phys.~Rev.~Lett.}

\newcommand{\comment}[1]{}

\title{
{\normalsize \hfill {\sf UTHEP-487}} \\
\vspace*{-2pt}
{\normalsize \hfill {\sf UTCCS-P-2}} \\
Non-perturbative renormalization of 
meson decay constants in quenched QCD
for a renormalization group improved gauge action
}

\author{
CP-PACS Collaboration: \\
K.~Ide\rlap,$^{\rm 1}$
S.~Aoki\rlap,$^{\rm 1}$
R.~Burkhalter\rlap,$^{\rm 1,2}$\footnote
	{Present address : Julius Baer Investment Fund Services Ltd., Freigutstrasse 12, 8010 Zurich, Switzerland}
M.~Fukugita\rlap,$^{\rm 3}$
S.~Hashimoto\rlap,$^{\rm 4}$
K.-I.~Ishikawa\rlap,$^{\rm 1,2}$\footnote
	{Present address : Department of Physics, Hiroshima University,Higashi-Hiroshima, Hiroshima 739-8526, Japan}
T.~Ishikawa\rlap,$^{\rm 2}$
N.~Ishizuka\rlap,$^{\rm 1,2}$
Y.~Iwasaki\rlap,$^{\rm 1,2}$
K.~Kanaya\rlap,$^{\rm 1}$
T.~Kaneko\rlap,$^{\rm 4}$
Y.~Kuramashi\rlap,$^{\rm 4}$\footnote
	{Present address:  Center for Computational Physics, University of Tsukuba, Tsukuba, Ibaraki 305-8577, Japan}
V.~Lesk\rlap,$^{\rm 2}$\footnote
	{Present address: Department of Biological Sciences, Imperial College, London SW7 2AZ, U.K.}
M.~Okawa\rlap,$^{\rm 5}$
Y.~Taniguchi\rlap,$^{\rm 1}$
T.~Umeda\rlap,$^{\rm 2}$\footnote
	{Present address : Yukawa Institute for Theoretical Physics, Kyoto University, Kyoto 606-8502, Japan}
A.~Ukawa\rlap,$^{\rm 1,2}$
T.~Yoshi\'e$^{\rm 1,2}$\\[4mm]
}

\address{
$^1$Institute of Physics, University of Tsukuba,
Tsukuba, Ibaraki 305-8571, Japan \\
$^2$Center for Computational Physics, University of Tsukuba,
Tsukuba, Ibaraki 305-8577, Japan \\
$^3$Institute for Cosmic Ray Research, University of Tokyo,
Tanashi, Tokyo 188-8502, Japan \\
$^4$High Energy Accelerator Research Organization (KEK),
Tsukuba, Ibaraki 305-0801, Japan \\
$^5$Department of Physics, Hiroshima University,
Higashi-Hiroshima, Hiroshima 739-8526, Japan \\[4mm]
}

\begin{document}
  \draft
  \tightenlines
  \maketitle

\begin{abstract}
Renormalization constants ($Z$-factors ) of vector and axial-vector
currents are determined non-perturbatively in quenched QCD for 
a renormalization group improved gauge action and a tadpole improved
clover quark action using the Schr\"odinger functional method.
Non-perturbative values of $Z$-factors turn out to be smaller than
one-loop perturbative values by $O(15\%)$
at lattice spacing of $a^{-1}\approx$ 1 GeV.
The pseudoscalar and vector meson decay constants calculated with 
the non-perturbative $Z$-factors show a much better scaling 
behavior compared to previous results obtained with tadpole improved
one-loop $Z$-factors.  In particular, the non-perturbative $Z$-factors 
normalized at infinite physical volume show that scaling violation 
of the decay constants are within about 10\% up to the lattice 
spacing $a^{-1}\sim 1$~GeV.  The continuum estimates obtained from 
data in the range $a^{-1}=$
1 -- 2 GeV agree with those determined from finer lattices 
($a^{-1}\sim 2-4$~GeV) with the standard action.   
\end{abstract}

\section{Introduction}
%%%% General Introduction %%%%
Reliable prediction of physical quantities from lattice 
QCD calculations requires a good control of scaling violations. 
For this purpose, several improved actions have been tested and 
applied to large scale systematic simulations.  
For most of physical quantities, 
such as quark masses and hadronic matrix elements, 
one, in addition, has to calculate renormalization constants 
($Z$-factors). Non-perturbative methods to determine various 
$Z$-factors have been developed and utilized for several actions.
% Compared to perturbative expansions of $Z$-factors,
% non-perturbative methods remove terms proportional 
% to powers of bare coupling, and hence improve further 
% scaling properties of physical quantities. 

%%%% Introduction of CP-PACS Action %%%%
%$Z$-factors should be determined action by action.
%For some cases, non-perturbative estimates of $Z$-factors 
%are missing for the action one currently employs for 
%systematic large scale calculations.
%For example, the CP-PACS collaboration carried out
%simulations for two-flavor full QCD~\cite{ref:CPPACS-NF2} using a
%renormalization group (RG) improved gauge action\cite{ref:Iwasaki}
%and a tadpole improved~\cite{ref:TP} clover quark action\cite{ref:Clover}
%and reported results for the spectrum, quark masses and meson decay
%constants.  
%Because non-perturbative $Z$-factors were not available 
%for this action combination, analysis had to rely on one-loop 
%perturbative values.

The CP-PACS collaboration carried out
systematic simulations for two-flavor full QCD~\cite{ref:CPPACS-NF2} using a
renormalization group (RG) improved gauge action\cite{ref:Iwasaki}
and a tadpole improved~\cite{ref:TP} clover quark action\cite{ref:Clover}
and reported results for the spectrum, quark masses and meson decay
constants.  
Since non-perturbative $Z$-factors were not available 
for this action combination, analysis had to rely on one-loop 
perturbative values.
The result showed that the meson decay constants 
in both quenched and full QCD suffer from large scaling violations 
at $a^{-1}\sim 1-2$~GeV which hinder continuum extrapolations.  
It was not clear if this observation hinted at an inherent difficulty
of improved actions for matrix elements at coarse lattice spacings or 
a perturbative treatment of the $Z$-factors was the issue. 
A non-perturbative determination of $Z$-factors was evidently needed. 

%%%% Purpose of this work %%%%
As a first step toward a systematic study of non-perturbative
renormalization for this action, we apply the Schr\"odinger functional 
(SF) method~\cite{ref:SFMethod,ref:ALPHA-SF,ref:ALPHA-mq,ref:ALPHA-Z}
to calculations of $Z$-factors for vector ($Z_V$) and 
axial-vector ($Z_A$) currents in quenched QCD 
with the same improved action. 
%For this action, the CP-PACS collaboration observed a large scaling 
%violation for meson decay constants calculated with perturbative 
%$Z$-factors in both quenched and full QCD 
%and did not draw any conclusions on values in the continuum limit.
%Therefore our primary concern is whether and how the large scaling
%violation is reduced when one employs non-perturbative $Z$-factors. 
% Precise continuum values of meson decay constants in quenched QCD 
% have already been determined by a large-scale simulation with 
% the plaquette gauge and the Wilson quark
% actions~\cite{ref:CPPACS-quench}. 
% Thus a calculation of $Z_V$ and $Z_A$ for the improved action is 
% a good test-bed to study how $Z$-factors affect scaling properties.
%%ukawa% no change of paragraph
%%%% Comment on feasibility study %%%%
The SF method has been applied to the non-perturbatively $O(a)$
improved Wilson action. 
On the other hand, our action combination has $O(ag^4)$ error, 
since the coefficient of the clover term is determined by a tadpole improved 
perturbation theory to one-loop order. 
%Therefore we investigate whether the SF method is successfully applied
%to this action.  
%%ukawa
Therefore our study involves an examination whether the SF method successfully
works for this action
%%ukawa
 
%%%% Z at finite L vs those at infinite L %%%% 
In the SF method, $Z$-factors are determined at various couplings
%for a fixed physical volume $L$. 
for a fixed physical size $L$.
In this case, the $Z$-factors contain terms of $O(ag^4/L)$
in addition to that of $O(a\Lambda)$.
We expect that one can remove the $O(ag^4/L)$ terms 
by taking the infinite volume limit.
We calculate $Z$-factors both at a finite fixed physical volume and 
at the infinite volume and compare scaling properties of 
decay constants with these two choices of $Z$-factors.

%%%% Exceptional configurations %%%%
When we calculate $Z$-factors for large couplings and on large
lattices, we encounter ``exceptional configurations'' for which 
observables take abnormally large  values.
We estimate systematic uncertainties due to such configurations, 
propagate them to systematic error estimations of decay constants, 
and argue that they do not alter 
our conclusions on scaling properties.

%%%% Paper organization %%%%
The organization of this paper is as follows. 
In Sec.\ref{sec:calc} we describe calculational method focusing
on features of applying the SF method to the RG improved action. 
Detail of analysis for $Z_V$ and $Z_A$ are given 
in Sec.~\ref{sec:renorm}. 
With these $Z$-factors, we study scaling behavior of decay constants
of vector ($f_\rho$) and pseudoscalar ($f_\pi$) mesons in
Sec.~\ref{sec:decay}. 
Sec.~\ref{sec:conclusion} is devoted to conclusions. 
Parts of this work have already been reported in
Refs.~\cite{ref:CPPACS-NPZ-RG-1} and ~\cite{ref:CPPACS-NPZ-RG-2}.
  
\section{Calculational Method}\label{sec:calc}

\subsection{Action and currents}
The RG improved gauge action we employ has the form
\begin{eqnarray}
S_g&=&\frac{\beta}{6}\left\{c_0\sum_{x,\mu,\nu}U_{P,\mu\nu}(x)
	+c_1\sum_{x,\mu,\nu}U_{R,\mu\nu}(x)\right\},
\end{eqnarray}
where $\beta={6}/{g^2}$ with the bare coupling constant $g$, and
$U_{P}$ ($U_{R}$) is
the trace of the product of link variables around a plaquette (6-link rectangular loop).
In a sum over loops, each oriented loop appears once.
The coefficients $c_1=-0.331$ and $c_0=1-8 c_1=3.648$ are fixed
by an approximate RG analysis~\cite{ref:Iwasaki}.
\comment{
the coefficients $c_1=-0.331$ and $c_0=1-8 c_1=3.648$ of the $1\times 1$ 
($W^{1\times 1}_{\mu\nu}$) and $1\times 2$ ($W^{1\times 2}_{\mu\nu}$)
which are the standard plaquette and rectangle terms,
are fixed by an approximate RG analysis~\cite{ref:Iwasaki}.
}

The clover quark action~\cite{ref:Clover} is defined by
\begin{eqnarray}
S_q&=&\sum_{x,y}\overline{\psi}(x)D_{x,y}\psi(y),\\
D_{x,y}&=&\delta_{x,y}
	-\kappa\sum_{\mu}\left\{(1-\gamma_\mu)U_{x,\mu}\delta_{x+\hat{\mu},y}
		+(1+\gamma_\mu)U_{x,\mu}^\dag\delta_{x,y+\hat{\mu}}\right\}
	-\delta_{x,y}c_{SW}\kappa\sum_{\mu<\nu}\sigma_{\mu\nu}F_{\mu\nu},
\end{eqnarray}
where $\kappa$ is the hopping parameter and $F_{\mu\nu}$ is
a lattice discretization of field strength. 
%For the clover coefficient $c_{SW}$, we adopt a mean field improved
For the clover coefficient $c_{SW}$, we adopt a tadpole improved
value with one-loop estimate of the plaquette
$\left<W^{1\times 1}\right>=\frac{1}{3}\left<U_{P,\mu\nu}\right>$ given by 
\begin{eqnarray}
c_{SW}&=&\left<W^{1\times 1}\right>^{-3/4}=(1-0.8412/\beta)^{-3/4},
\end{eqnarray}
since the one-loop value of the plaquette 
reproduces measured values well. 
%$\left<U_P\right>=\frac{1}{3}\left<W^{1\times 1}\right>$ is the expectation value of plaquette.

% Our action combination has $O(ag^4)$ error, because $c_{SW}$ is tuned
% by tadpole improved perturbation theory. The SF method utilizes Ward
% identities at zero quark mass~\cite{ref:ALPHA-Z} so that the physical
% volume $L$ is the unique scale. For our case, resulting $Z$-factors
% has systematic error of $O(ag^4/L)$. We estimate magnitude of the
% error in Sec.\ref{sec:renorm} by varying $L$. 

For $Z_V$, we investigate vector Ward identity of 
the unimproved current $V_\mu^a$ since the particular SF setup we use yields 
the same value of $Z_V$ for improved and unimproved currents.
On the other hand, we study an improved current for $Z_A$ defined by 
\begin{equation}
(A_I)_\mu^a(x) =
 A_\mu^a(x)+a c_A\frac{1}{2}(\partial_\mu^*+\partial_\mu)P^a(x),
\end{equation}
where $\partial_\mu^*$ and $\partial_\mu$ are lattice forward and 
backward derivatives and $P^a(x)$ is the pseudoscalar density.
We use one-loop value for the improvement coefficient 
$c_A=-0.0038 g^2$~\cite{ref:CA-RG}.

\subsection{Implementation for RG improved action} 
We follow the SF method of Ref~\cite{ref:ALPHA-Z} for the RG improved
gauge action. 
% The only difference resides in the choice of the action and parameters. 
Taking an $L^3\times T$ lattice, we impose the periodic
boundary condition in the spatial directions and the Dirichlet boundary
condition in the temporal direction.

The boundary counter terms for the RG improved action are 
determined~\cite{ref:SF-RG} so that the classical field equations are satisfied.
The total action reads
\begin{eqnarray}
S_g&=&\frac{\beta}{6}\sum_{\bf x}\sum_{x_4=0}^{T}
	\left\{c_0\sum_{\mu,\nu}
		w^{P}_{\mu\nu}(x_4) U_{P,\mu\nu}(x)
	+c_1\sum_{\mu,\nu}
		w^{R}_{\mu\nu}(x_4) U_{R,\mu\nu}(x)\right\},
\end{eqnarray}
with the weight factors $w^{P}_{\mu\nu}(x_4)$ and 
$w^{R}_{\mu\nu}(x_4)$ given by
\begin{eqnarray}
w^{P}_{\mu\nu}(x_4)&=&\left\{\begin{array}{cl}
		c_t            &
		\mbox{($x_4=0$ or $x_4=T-a$) and ($\mu=4$ or $\nu=4$)}\\
		0              &
		\mbox{$x_4=T$ and ($\mu=4$ or $\nu=4$)}\\
		\frac{1}{2}c_s &
		\mbox{($x_4=0$ or $x_4=T$) and ($\mu\neq 4$ and $\nu\neq 4$)}\\
		1              &
		\mbox{otherwise.}
	\end{array}\right.,\\
w^{R}_{\mu\nu}(x_4)&=&\left\{\begin{array}{cl}
		\frac{3}{2}&\mbox{when 2 links of the rectangular touch a boundary}\\
		0&\mbox{when the rectangular is completely included in 
		a boundary}\\
		1&\mbox{otherwise.}
	\end{array}\right.
\end{eqnarray}
We take the tree-level value of $c_s=c_t=1$.
(One-loop values of $c_s$ and $c_t$~\cite{ref:CT-RG} were not known
when we started this work.)
Boundary counter terms are not included for the clover quark action.
In other words, the boundary coefficients $\tilde{c}_s$ and
$\tilde{c}_t$ of Ref.~\cite{ref:ALPHA-Z} are set to the tree-level values,  
$\tilde{c}_s=\tilde{c}_t=1$. 

\subsection{Details of calculational method}

%%%% Simulation parameters %%%%
We calculate $Z$-factors for the range $\beta=8.0 - 2.2$.
The smallest value $\beta=2.2$ is chosen so as to cover 
the range $\beta=2.575 - 2.184$ (inverse lattice spacing 
$a^{-1} \approx 1-2$ GeV) where data for $f_\rho$ and $f_\pi$ 
exist~\cite{ref:CPPACS-NF2}. 
Simulations are made with a 5-hit pseudo-heat-bath algorithm 
mixed with an over-relaxation algorithm in the ratio of 1:4. 
We analyze 200 -- 20000 configurations separated by 100 sweeps each.
Lattice geometry is set to $T=2L$ for both $Z_V$ and $Z_A$.
% though $T=2L$ for $Z_V$ and $T=\frac{9}{4}L$ for $Z_A$ were employed 
% in Ref.~\cite{ref:ALPHA-Z}.
% This choice was made to suppress ``exceptional configurations''
% at small $\beta$'s. 
% See Sec.~\ref{subsec:except} for details.
At each $\beta$, $Z$-factors are determined for at least two lattice
sizes in order to interpolate or extrapolate them to a fixed physical 
volume.
Lattice size and number of analyzed configurations are listed
in Table~\ref{tab:kcmqZVZAatSP}.

%%%% Critical hopping parameters %%%%
Calculations of $Z$-factors are carried out at zero quark mass, where
the quark mass $m_q$ is defined by the PCAC relation 
$\partial_\mu A_\mu^a = 2m_q P^a$ in the continuum notation. 
The 
actual procedure to measure $m_q$ is the same as in Ref.~\cite{ref:ALPHA-mq}. 
We define a time-dependent quark mass by 
\begin{equation}
m_q(x_4)=\frac{\frac{1}{2}\left(\partial_0^*+\partial_0\right)f_A(x_4)
		+c_Aa\partial_0^*\partial_0f_P(x_4)}
	{2f_P(x_4)}, \label{eq:AWI-x4}
\end{equation}
where $f_A$ is the axial-vector and pseudoscalar correlator and 
$f_P$ is the pseudoscalar and pseudoscalar correlator.
Quark mass $m_q$ is defined by an average of $m_q(x_4)$ over a range of time
slice around $x_4 = T/2$;
\begin{equation}
m_q=\frac{1}{2n+1}\sum_{t=-an}^{an}m_q\left(\frac{T}{2}+t\right),
	\label{eq:AWI-mass}
\end{equation}
where $n$ defines the range and depends on simulation parameters.

%%% Z-factors %%%%
$Z$-factors are determined from correlators of pseudoscalar operators 
$P^a$ and $P'^a$ at the boundaries and/or currents
$V_\mu^a$ and $(A_I)_\mu^a$ at a finite time slice. 
We employ the notations defined by 
\begin{eqnarray}
f_1&=&-\frac{1}{3L^6}\left< P'^a  P^a \right>,\label{eq:f1} \\
f_V(x_4)&=& \frac{a^3}{6L^6}\sum_{\bf x}i\epsilon^{abc}
	\left<P'^aV_0^b(x)P^c\right>, \label{eq:fV}\\
f_{AA}(x_4,y_4)&=&-\frac{a^6}{6L^6}\sum_{{\bf x},{\bf y}}
	\epsilon^{abc}\epsilon^{cde}\left<P'^d
	(A_I)_0^a(x)(A_I)_0^b(y)P^e\right>. \label{eq:fAA}
\end{eqnarray}
Renormalization constants are then extracted from 
\begin{eqnarray}
Z_V&=&\frac{f_1}{f_V(\frac{T}{2})},\\
Z_A&=&\sqrt{\frac{f_1}{f_{AA}(\frac{3T}{8},\frac{5T}{8})}}.
\end{eqnarray}
% after we check the existence of plateau of the ratios.  

%%%% Physical volume %%%%
Lattice spacing necessary to set the physical size is determined
from the string tension $\sqrt\sigma=440$ MeV.
We fit values of $a\sqrt\sigma$~\cite{ref:CPPACS-NF2,ref:Topology}
to a fitting form~\cite{ref:Allton}   
\begin{eqnarray}
(a\sqrt{\sigma})(\beta)&=&f(\beta)(1+c_2\hat{a}^2(\beta)+c_4\hat{a}^4(\beta)
	+\cdots)/c_0,
	\quad \hat{a}(\beta)\equiv\frac{f(\beta)}{f(\beta_1)},
\end{eqnarray}
where $f(\beta)$ is the two-loop scaling function of the SU(3) gauge
theory, $c_n$'s are parameters to describe deviation from the
two-loop scaling, and $\beta_1$ is a reference point. 
Choosing $\beta_1=2.4$, the parameters
\begin{equation}
c_0=0.5443(97),\quad c_2=0.390(38),\quad c_4=0.049(12)
\end{equation}
reproduce measured values well.

\section{Renormalization Constants}\label{sec:renorm}
\subsection{Results at simulation points}
%%%% Focus on normal cases %%%%
Calculations of $Z$-factors do not present any difficulty from $\beta=8.0$ down
to 2.6 for all lattice sizes,  and at $\beta=2.4$ and 2.2
on a small lattice of $4^3\times 8$.
For a larger $8^3\times 16$ lattice at $\beta=2.4$ and 2.2, however, 
we encounter the issue of ``exceptional configurations''. 
Deferring discussions of this issue, 
let us first summarize results for the non-exceptional case.

\begin{figure}[htbp]
%\centerline{\epsfysize=10.0cm \epsfbox{eps/k-2.6-8x16-rev.eps}}
\centerline{\epsfysize=10.0cm \epsfbox{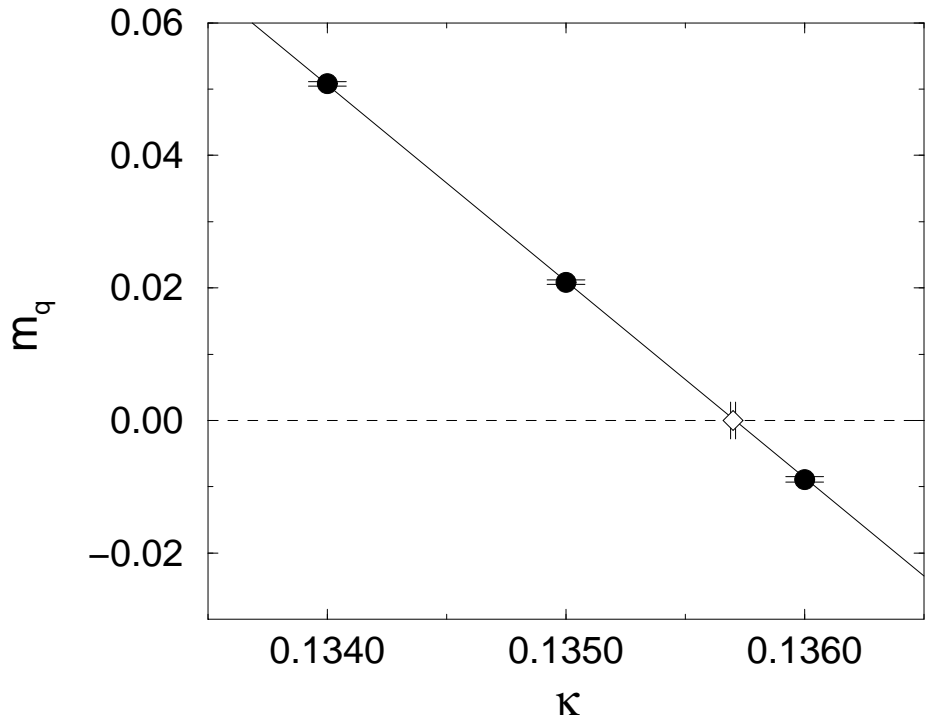}}
\vspace{0.5cm}
\caption{$m_q$ versus $\kappa$ at $\beta=2.6$ on an $8^3\times 16$
lattice. Filled symbols show measured points. 
The open symbol represents $\kappa_c$.}
\label{fig:kappa-b2.6-8x16}
\end{figure}

%%%% Critical hopping parameters %%%%
We first determine $\kappa_c$ for each $\beta$ and $L/a$.
For this purpose simulations are carried out at several values of 
$\kappa$ around an estimated $\kappa_c$ in which 
$m_q$ is determined by eq.(\ref{eq:AWI-mass}); we employ the   
fitting range given by $n=$ 0 for $L/a=4$ lattices, $n=1$ for $L/a=8$
lattices at $\beta=2.4$ and 2.2, and $n=2$ for $L/a=8$ at other values of
$\beta$ and $L/a=16$ lattices.
We determine $\kappa_c$ by a linear fit in $\kappa$ 
as illustrated in Fig.~\ref{fig:kappa-b2.6-8x16} 
for an $8^3\times 16$ lattice at $\beta=2.6$. 

\begin{figure}[htbp]
%\centerline{\epsfysize=10.0cm \epsfbox{eps/t-m-2.6-8x16-rev.eps}}
\centerline{\epsfysize=10.0cm \epsfbox{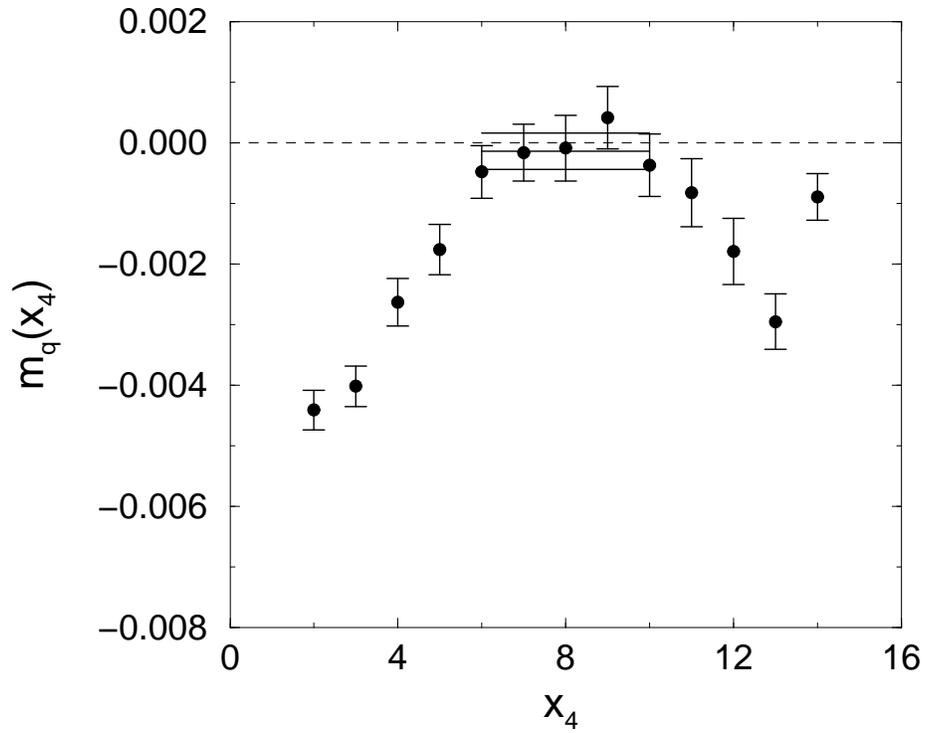}}
\vspace{0.5cm}
\caption{Effective quark mass $m_q(x_4)$ at $\beta=2.6$ 
on an $8^3\times 16$ lattice. 
Horizontal solid lines represent fitting range, and fitted value and error.}
\label{fig:mqeff-b2.6-8x16}
%\vspace{-0.7cm}
\end{figure}

Results for $\kappa_c$ are given in Table~\ref{tab:kcmqZVZAatSP}.  
When we perform simulations for $Z$-factors at $\kappa_c$, 
$m_q$ is also calculated for confirmation. 
Typical result for the effective quark mass $m_q(x_4)$ is shown in 
Fig.~\ref{fig:mqeff-b2.6-8x16}. 
It exhibits a reasonable plateau.
Values of $m_q$ at $\kappa_c$ are given in the fourth column of
Table~\ref{tab:kcmqZVZAatSP}.
They are consistent with zero within an accuracy of $10^{-3}$ at worst.
Since $\kappa_c$ is tuned well, errors of $\kappa_c$ are not taken
into account in the error estimation of $Z$-factors. 

\begin{figure}[htbp]
%\centerline{\epsfysize=10.0cm \epsfbox{eps/t-Zv-16x32-3.125-rev.eps}}
\centerline{\epsfysize=10.0cm \epsfbox{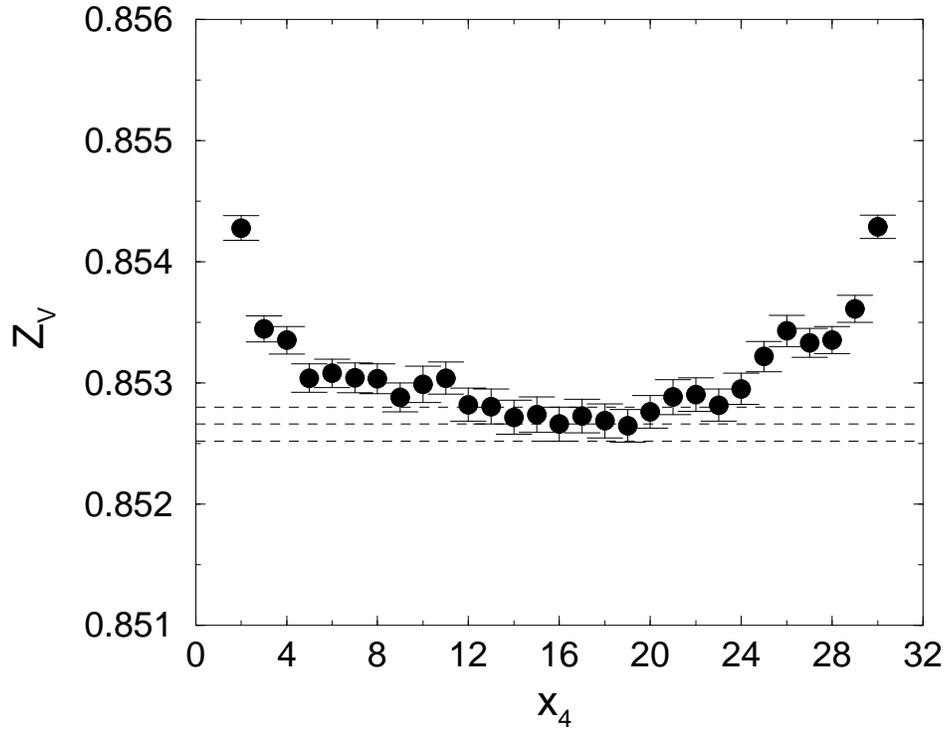}}
%\vspace{-1.1cm}
\caption{$Z_V$ as a function of time slice
at $\beta=3.125$ on a $16^3\times 32$ lattice. Horizontal dashed
lines represent $Z_V$ at $x_4=T/2=16$.}
\label{fig:t-ZV}
%\vspace{-0.7cm}
\end{figure}

\begin{figure}[htbp]
%\centerline{\epsfysize=10.0cm \epsfbox{eps/t-ZA-16x32-3.125.eps}}
\centerline{\epsfysize=10.0cm \epsfbox{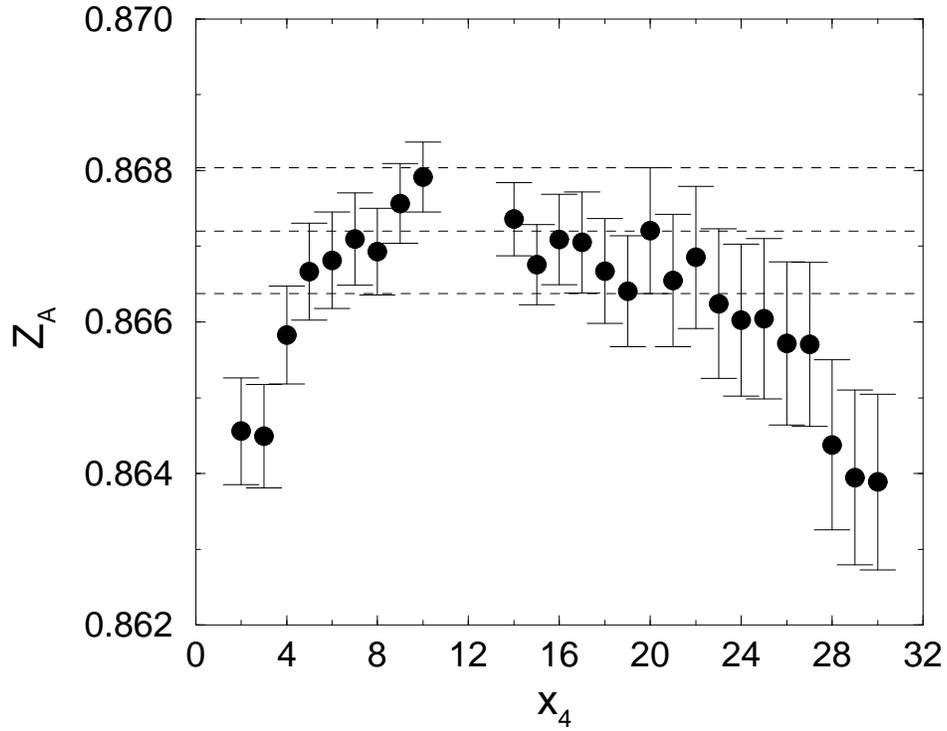}}
%\vspace{-1.1cm}
\caption{Values of $Z_A$ as a function of time slice
at $\beta=3.125$ on a $16^3\times 32$ lattice. Horizontal dashed
lines represent $Z_A$ at $x_4=(5/8)\cdot T=20$.}
\label{fig:t-ZA}
%\vspace{-0.7cm}
\end{figure}

%%%% Plateau of ZV and Z_A %%%%
Figures~\ref{fig:t-ZV} and ~\ref{fig:t-ZA} show typical plots of 
$Z_V$ and $Z_A$ as a function of time slice. 
We observe good plateaux for all cases, except for a small temporal lattice 
size of $L/a=8$ at strong coupling of $\beta=2.4$ and 2.2. 
We thus find that the SF method works successfully to 
our action, albeit there are $O(ag^4)$ errors rather than $O(a^2)$ errors.
Results for $Z$-factors at simulation points are given in 
Table~\ref{tab:kcmqZVZAatSP}.

\begin{figure}[htbp]
%\centerline{\epsfysize=10.0cm \epsfbox{eps/fAfp-history-2.4-8x16-rev.eps}}
\centerline{\epsfysize=10.0cm \epsfbox{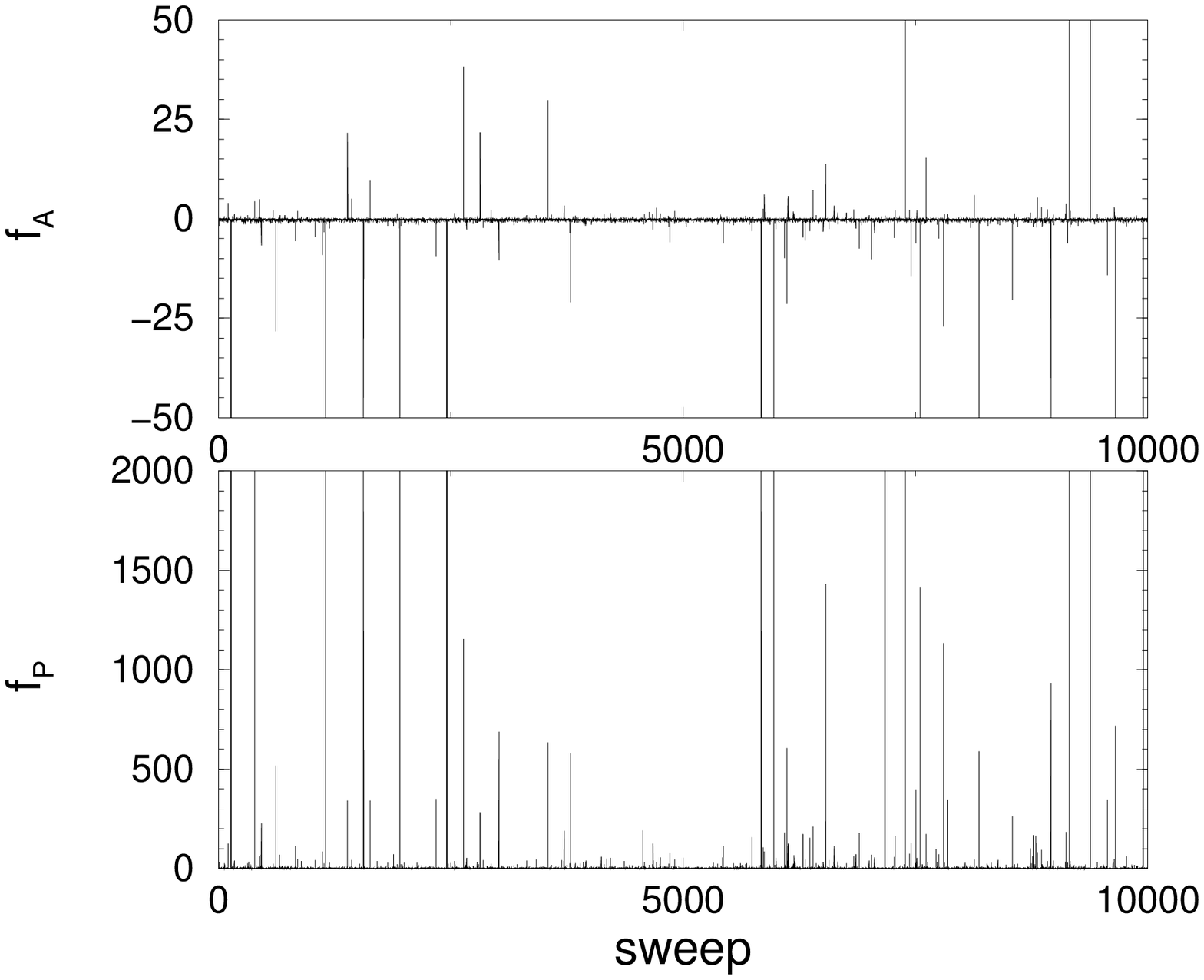}}
\vspace{0.5cm}
\caption{Time history of $f_A$ and $f_P$ at $\beta=2.4$ 
on an $8^3\times 16$ lattice.}
\label{fig:fAfp-hist-2.4-8x16}
%\vspace{-0.7cm}
\end{figure}

\begin{figure}[htbp]
%\centerline{\epsfysize=10.0cm \epsfbox{eps/fAfp-history-2.6-8x16.eps}}
\centerline{\epsfysize=10.0cm \epsfbox{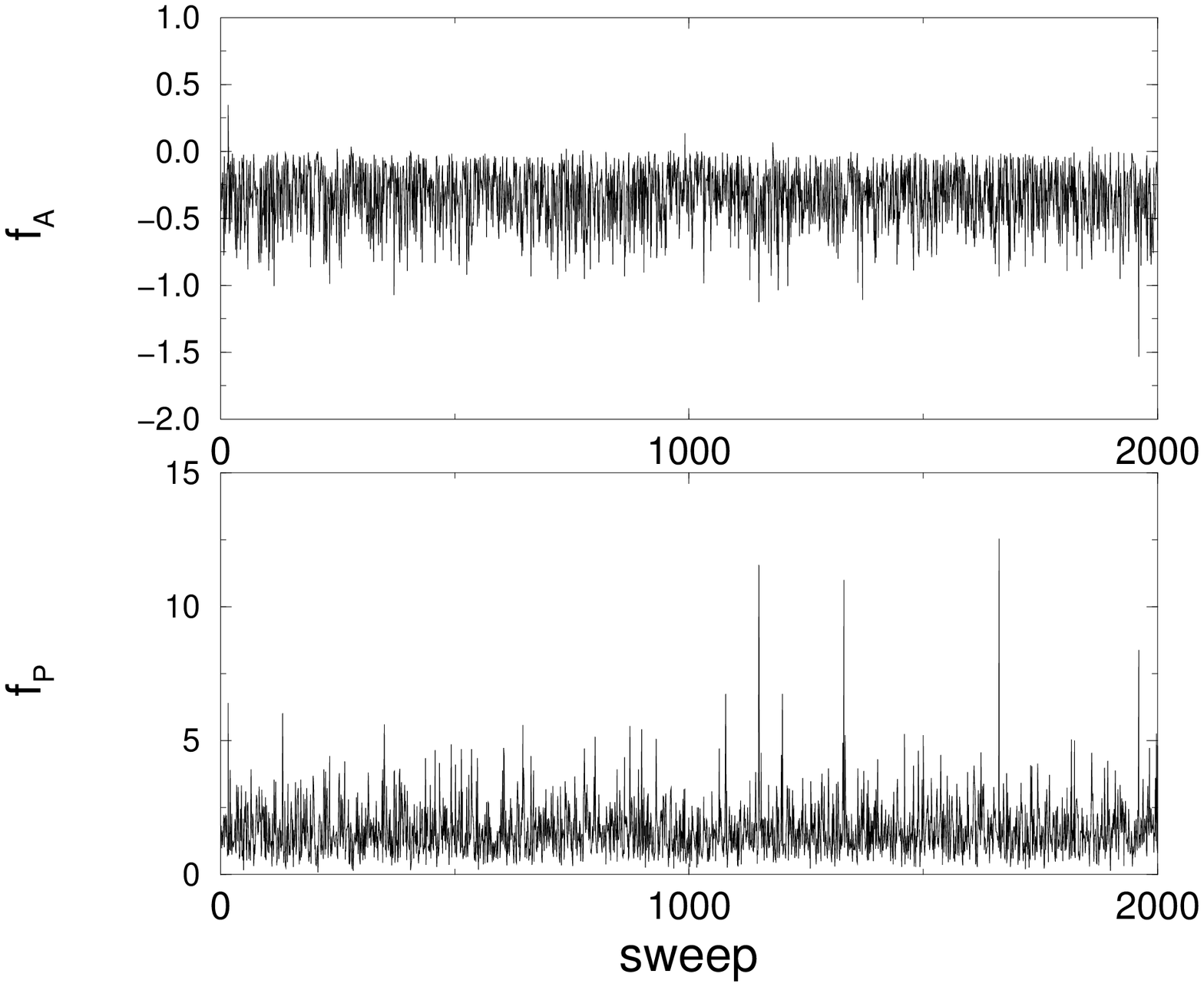}}
\vspace{0.5cm}
\caption{Time history of $f_A$ and $f_P$ at $\beta=2.6$ on an $8^3\times 16$ lattice.
}
\label{fig:fAfp-hist-2.6-8x16}
%\vspace{-0.7cm}
\end{figure}

%%%% Exceptional configurations %%%%
We now discuss the issue of ``exceptional configurations''. 
Among our parameter sets, anomalously large values appear
in the ensemble of hadron correlators on an $8^3\times 16$ lattice 
at $\beta=2.4$ and 2.2.
We illustrate the situation in Fig.~\ref{fig:fAfp-hist-2.4-8x16}, where
we plot the time history of $f_A$ and $f_P$ at $\beta=2.4$.
No large spikes appear for larger $\beta$ on the same size lattice, 
as shown in Fig.~\ref{fig:fAfp-hist-2.6-8x16} for data at $\beta=2.6$.
Note that a much finer vertical scale is employed in the latter
figure. 
%We call configurations which shows large spikes as ``exceptional''.
Large spikes appear also in the ensembles of $f_1$, $f_V$ and $f_{AA}$
necessary for evaluating $Z$-factors.
Such Exceptional configurations make it difficult to determine $m_q$,
$\kappa_c$ and $Z$-factors precisely.

\begin{figure}[htbp]
%\centerline{\epsfysize=10.0cm \epsfbox{eps/fp-histogram-2.4-8x16.eps}}
\centerline{\epsfysize=10.0cm \epsfbox{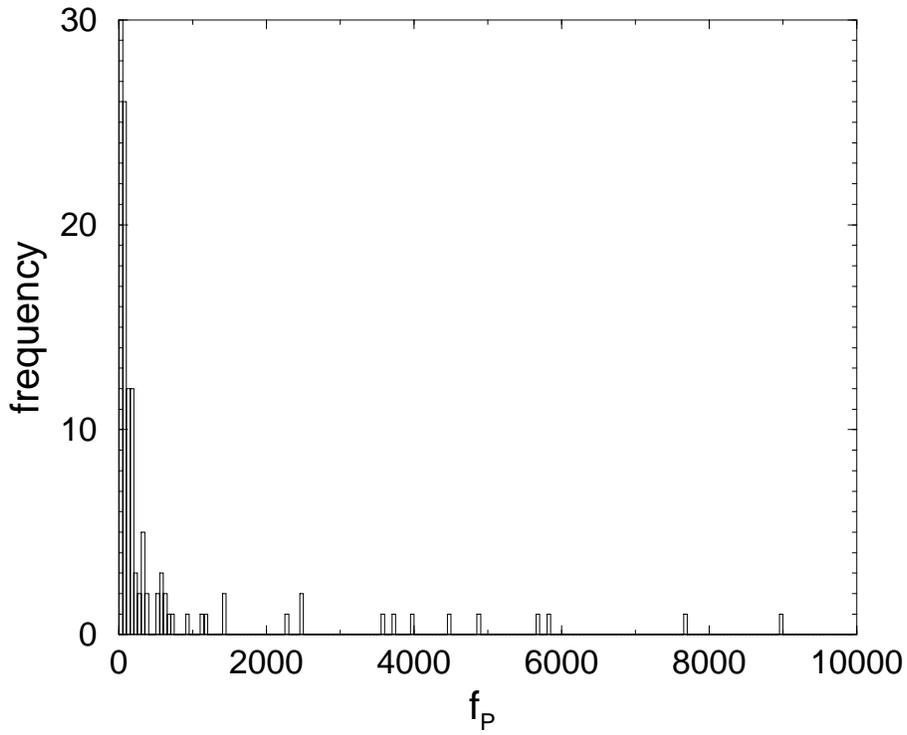}}
\vspace{0.5cm}
\caption{Histogram of $f_P$ at $\beta=2.4$ on an $8^3\times 16$ lattice.
Intervals in $x$-axis are 50.}
\label{fig:fp-histogram-2.4-8x16}
%\vspace{-0.7cm}
\end{figure}

%%%% Strategy %%%%
We suspect that the spikes are caused by the appearance of 
very small or even negative eigenvalues of the Dirac operator toward 
strong coupling.  Such eigenvalues would be suppressed in full QCD by the 
quark determinant, and in this sense we expect exceptional configurations 
to be an artifact of quenched QCD.
In quenched QCD, however,  ``exceptional configurations'' cannot be 
distinguished from ``normal'' ones on some rigorous basis. In fact, 
histograms of $f$'s have a long tail 
toward very large values as shown in Fig.~\ref{fig:fp-histogram-2.4-8x16}.
We then adopt the strategy of estimating $Z$-factors by removing from 
the ensemble average configurations having values above some cutoff.
The uncertainties associated with this procedure are estimated by 
varying the cutoff, and will be
propagated to systematic errors of $Z$-factors at fixed physical volume.
Detailed description of this procedure will be given in 
Sec.~\ref{subsec:except}.

\subsection{$Z$-factors at fixed physical volume}\label{subsec:resZ}

\begin{figure}[htbp]
%\centerline{\epsfysize=10.0cm \epsfbox{eps/N_inv-Zv-Oa.eps}}
\centerline{\epsfysize=10.0cm \epsfbox{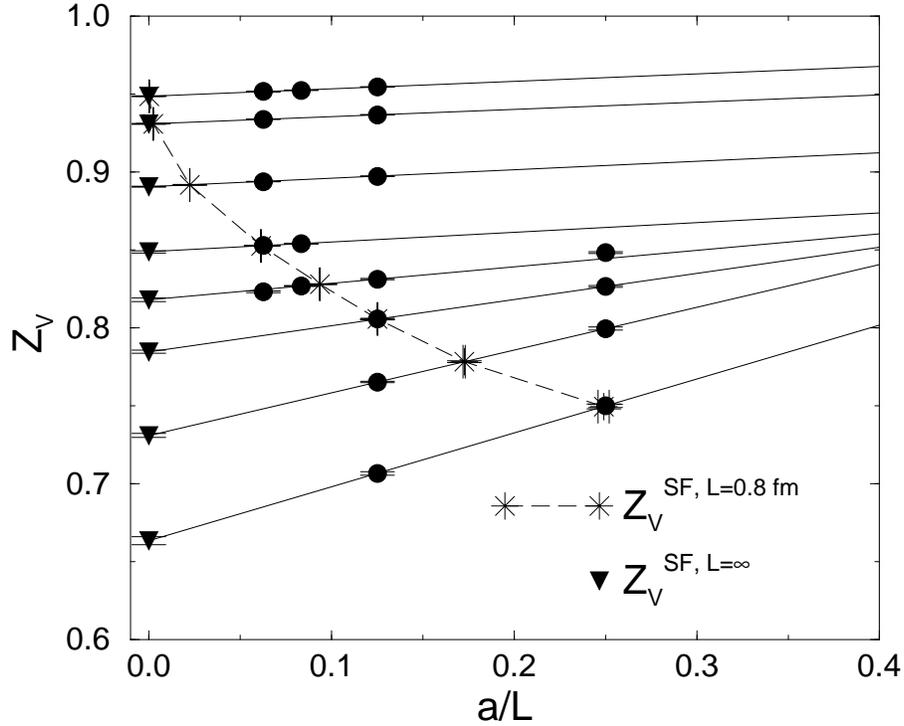}}
\vspace{0.5cm}
\caption{Volume dependence of $Z_V$. 
Filled circles and solid lines are measured data and linear fits to them. 
Values of $\beta$ are, from top to 
bottom, 8.0, 6.0, 4.0, 3.125, 2.8, 2.6, 2.4 and 2.2, respectively.
Stars $Z_V^{SF,L=0.8\rm{ fm} }$ are connected by dashed lines
to guide eyes. Filled triangles at $a/L=0$ are $Z_V^{SF,L=\infty}$.} 
\label{fig:ZV-vs-1L}
%\vspace{-0.7cm}
\end{figure}

\begin{figure}[htbp]
%\centerline{\epsfysize=10.0cm \epsfbox{eps/N_inv-ZA-Oa.eps}}
\centerline{\epsfysize=10.0cm \epsfbox{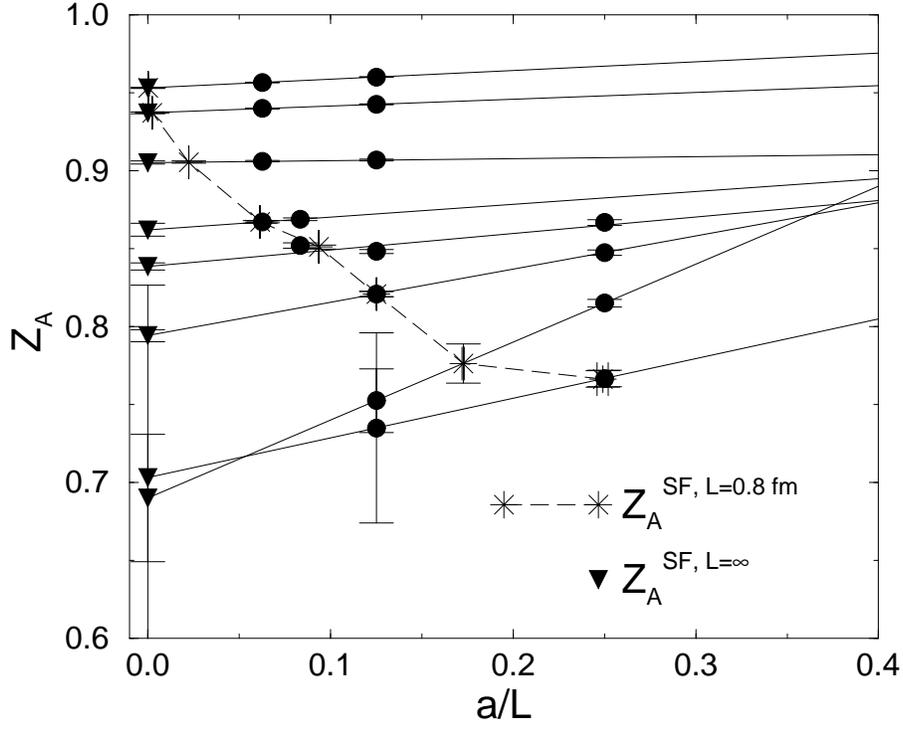}}
\vspace{0.5cm}
\caption{Volume dependence of $Z_A$. Symbols are the same as
in Fig.~\protect\ref{fig:ZV-vs-1L}. }
\label{fig:ZA-vs-1L}
%\vspace{-0.7cm}
\end{figure}

%%%% Linear fit for $Z$-factors at fixed physical volume %%%%
We plot $Z_V$ and $Z_A$ determined for various sizes and $\beta$ 
in Figs.~\ref{fig:ZV-vs-1L} and \ref{fig:ZA-vs-1L},
respectively, as a function of $a/L$. 
There are three or four data points at $\beta=8.0$ and 2.8 for $Z_V$,
and at $\beta=2.8$ for $Z_A$.
These data show a linear behavior in $a/L$, which is consistent
with the expectation that $Z$-factors for our action
have $O(ag^4/L)$ errors. 
Therefore we adopt a linear ansatz to extrapolate or interpolate $Z$-factors 
to the physical volume of $L=0.8$~fm (normalized at $\beta=2.6$ 
and $L/a=8$) and to $L=\infty$, as shown in these figures.
We denote $Z$-factors at a fixed physical volume as
$Z_{V,A}^{SF,L=0.8\rm{ fm} }$ and $Z_{V,A}^{SF,L=\infty}$.
Numerical values are given in Table~\ref{tab:ZVZA-phys}.

\begin{figure}[htbp]
%\centerline{\epsfysize=10.0cm \epsfbox{eps/g2-Zv-Oa.eps}}
\centerline{\epsfysize=10.0cm \epsfbox{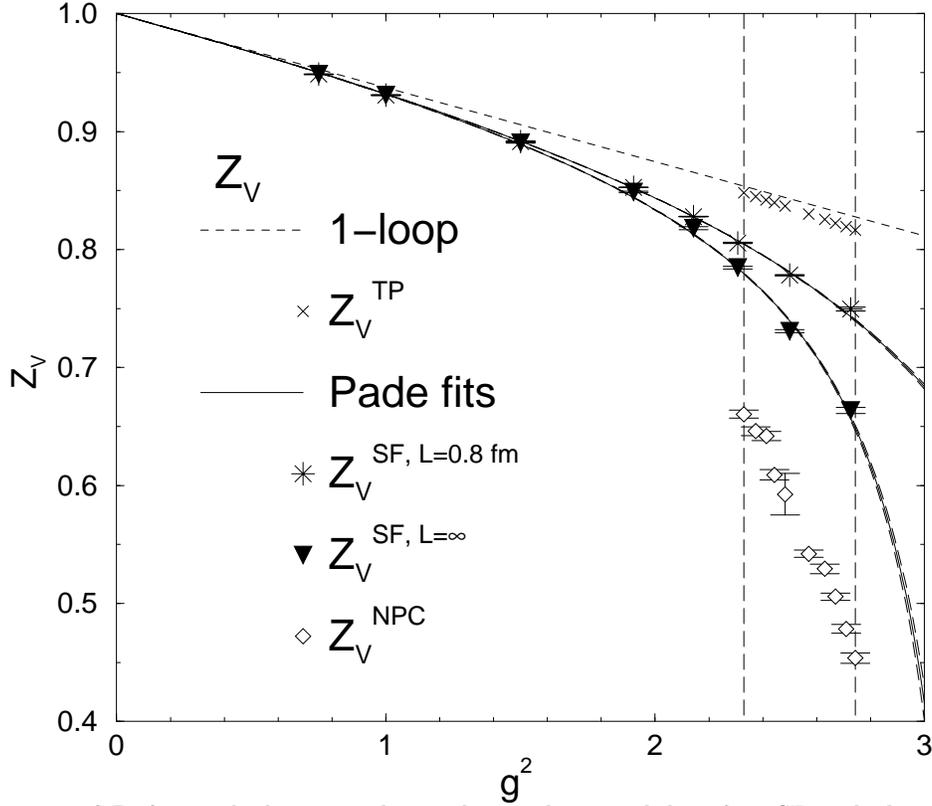}}
\vspace{0.5cm}
\caption{Comparison of $Z_V$ from tadpole improved perturbation
theory and those from SF method normalized at $L=0.8$~fm and $L=\infty$.
$Z_V^{\rm NPC}$ are determined with the conserved current. 
Pad\'e fits and error are also given for $Z_V$ from the SF method.
Vertical dashed lines represent the range we have data for decay
constants.}
\label{fig:ZV-fixed-L}
%\vspace{-0.7cm}
\end{figure}

%%%% results for $Z_V$ %%%%
In Fig.~\ref{fig:ZV-fixed-L} we plot $Z_V$ as a function of $g^2$.
Making a Pad\'e fit, we obtain 
\begin{eqnarray}
Z_V^{SF,L=0.8\rm{ fm} }&=&\frac{1-0.302225g^2+0.011034g^4}{1-0.239431g^2}, 
\label{eq:PadeZV08}\\
Z_V^{SF,L=\infty}&=&\frac{1-0.365802g^2+0.015016g^4}{1-0.303008g^2},
\label{eq:PadeZVinf} 
\end{eqnarray} 
where we have imposed a constraint that the Pad\'e fits reproduce
the one-loop perturbative result $Z_V= 1 - 0.062794 g^2$~\cite{ref:PTZ-RG} 
up to $O(g^2)$.

%%%%%%%% Comparison with perturbation theory %%%%%%%%
The range of coupling where there are data for the vector meson decay 
constant is marked by the two vertical dashed lines 
in Fig.~\ref{fig:ZV-fixed-L}.  Over this range, 
$Z_V^{SF,L=\infty}$ becomes increasingly smaller (by about 8 -- 21\%)  
compared to the one-loop value (dashed line) and 
the tadpole-improved value (crosses).
We also observe that $Z_V$ exhibits a sizable volume dependence 
toward strong coupling, {\it e.g.,} for $\beta\le 2.8$. 
This will have an important consequence on the scaling property of $f_\rho$
as discussed in Sec.~\ref{sec:decay}.

%%%%%%%% Comparison with NPC %%%%%%%%
Another method to estimate $Z_V$ non-perturbatively~\cite{ref:NPC}  
utilizes the non-renormalizability of the conserved vector
current 
\begin{equation}
V_i^C(x)=\frac{1}{2}\left\{
		\overline{\psi}(x+a\hat{\mu})
			U_{x,\mu}^\dag\left(\gamma_i+1\right)\psi(x)
		+\overline{\psi}(x)
			U_{x,\mu}\left(\gamma_i-1\right)\psi(x+a\hat{\mu})
	\right\}, \label{eq:VC}
\end{equation}
and defines 
\begin{equation}
%Z_V^{\mbox{NPC}}=
Z_V^{\rm NPC}=
\lim_{x_4\to\infty}\frac{\sum_{\bf x}\left<0\left|V_i^C({\bf x},x_4)V_i\right|0\right>}
     {\sum_{\bf x}\left<0\left|V_i({\bf x},x_4)V_i\right|0\right>}. \label{eq:ZNPC}
\end{equation}
%Results for $Z_V^{\mbox{NPC}}$ 
Results for $Z_V^{\rm NPC}$
obtained for our action combination~\cite{ref:CPPACS-NF2} are also  
overlaid in Fig.~\ref{fig:ZV-fixed-L} (open diamonds). 
They are much smaller than those from the SF method. 
We interpret that the large difference originates from large $O(a)$
%errors in $Z_V^{\mbox{NPC}}$.
errors in $Z_V^{\rm NPC}$.
From  the viewpoint of $O(a)$ improvement of operators, 
divergence of a tensor operator $\partial_\mu T_{i\mu}$ should
be added to both the local current $V_i$ and the conserved 
current $V_i^C$~\cite{ref:NPC-Oa}.
%However, this term $\sum_j \partial_j T_{0j}$ automatically drops 
%out in the SF scheme since it is a spatial total divergence.
However, the improvement operator $\sum_j \partial_j T_{0j}$, necessary for $V_0$ in the SF scheme,
automatically drops out since it is a spatial total divergence.
In other words, $Z_V$ from the SF method is $O(a)$ improved, whereas 
%the conserved current is not. 
%$Z_V^{\mbox{NPC}}$ is not.
$Z_V^{\rm NPC}$ is not.

\begin{figure}[htbp]
%\centerline{\epsfysize=10.0cm \epsfbox{eps/g2-ZA-Oa.eps}}
\centerline{\epsfysize=10.0cm \epsfbox{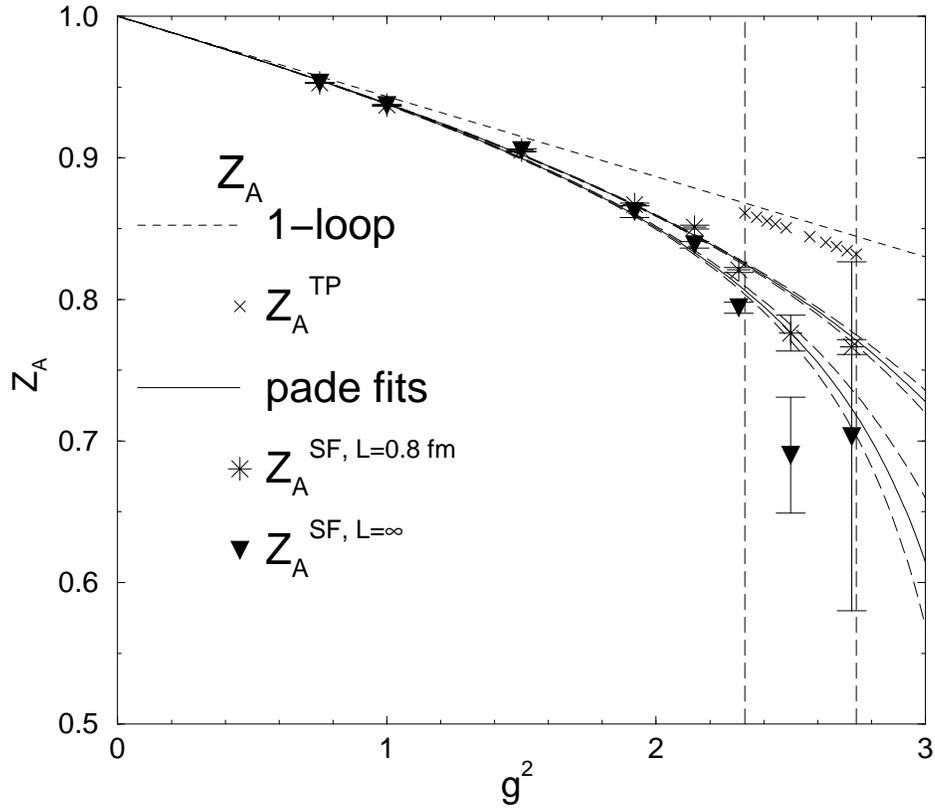}}
\vspace{0.5cm}
\caption{Comparison of $Z_A$ determined by various methods. 
Symbols are the same as in Fig.~\protect\ref{fig:ZV-fixed-L}.}
\label{fig:ZA-fixed-L}
%\vspace{-0.7cm}
\end{figure}

%%%% Z_A %%%%
In Fig.~\ref{fig:ZA-fixed-L}, we plot results for $Z_A$ and Pad\'e fits 
which read
\begin{eqnarray}
Z_A^{SF,L=0.8\rm{ fm} }&=&\frac{1-0.277576g^2+0.008669g^4}{1-0.220946g^2},\\
Z_A^{SF,L=\infty}&=&\frac{1-0.334232g^2+0.011710g^4}{1-0.277602g^2}.
\label{eq:ZAPade}
\end{eqnarray}
In the Pad\'e fits, we use a constraint from one-loop perturbation
theory that $Z_A=1 - 0.056630 g^2+O(g^4)$~\cite{ref:PTZ-RG}.
We observe that $Z_A^{SF,L=\infty}$ for the range we have data for 
$f_\pi$ are smaller than the tadpole improved value by 6 -- 14\%.

\subsection{Systematic error from exceptional configurations}\label{subsec:except}

Exceptional configurations affect $Z$-factors in two way, 
firstly by changing the value of $m_q$ and hence that of $\kappa_c$, 
and secondly by directly affecting the value of $Z$-factors themselves. 
Hence, in order to estimate the uncertainties of $Z$-factors 
due to exceptional configurations at large couplings, we investigate
how $m_q$ and $Z$-factors change if we discard configurations having
values of relevant correlators larger than some cutoff.

\begin{figure}[htbp]
%\centerline{\epsfysize=10.0cm \epsfbox{eps/mass-cutoff-dep-fp-2.4-8x16.eps}}
\centerline{\epsfysize=10.0cm \epsfbox{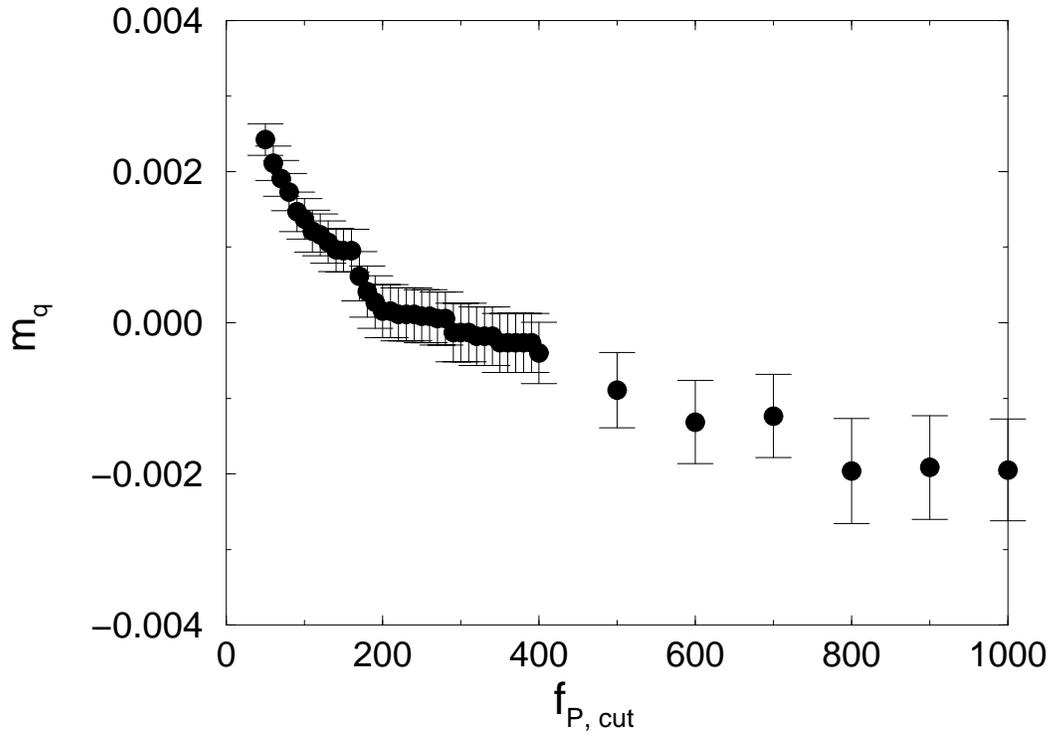}}
\vspace{0.5cm}
\caption{$m_q$ vs. $f_{P,{\rm cut}}$ at $\kappa_c$ 
on an $8^3\times 16$ lattice at $\beta=2.4$.}
\label{fig:mqvsfpat24}
\end{figure}

%%%% kappa_c  %%%%
Figure~\ref{fig:mqvsfpat24} shows this test for $m_q$ at $\beta=2.4$
for which a cutoff is set for the value of $f_P$. 
As the cutoff $f_{P,\mbox{cut}}$ is increased, 
$m_q$ gradually decreases, becomes almost stable around
$f_{P,\mbox{cut}}$=300 and then the error of $m_q$ becomes large.
With this observation in mind, we have estimated $\kappa_c$ as the 
point where $m_q$ for $f_{P,\mbox{cut}}$=300 is consistent with zero,
as shown in Fig.~\ref{fig:mqvsfpat24}.
The uncertainty in $m_q$ at $\kappa_c$ is estimated by varying 
$f_{P,\mbox{cut}}$ from 200 to 1000 and turns out to be 
$+0.00028 > m_q > -0.00182$. 
The same procedure at $\beta=2.2$ with $f_{P,\mbox{cut}}$=500 
gives the uncertainty $+0.00344 > m_q > -0.00170$. 
We note that the number of configurations discarded is 37 (191) of
the total of 10000 (20000) configurations at $\beta= 2.4$ (2.2). 
% 20 (117) for cut=1000 at b=2.4 (2.2).
 
\begin{figure}[htbp]
%\centerline{\epsfysize=10.0cm \epsfbox{eps/m-Zv-2.4-8x16.eps}}
\centerline{\epsfysize=10.0cm \epsfbox{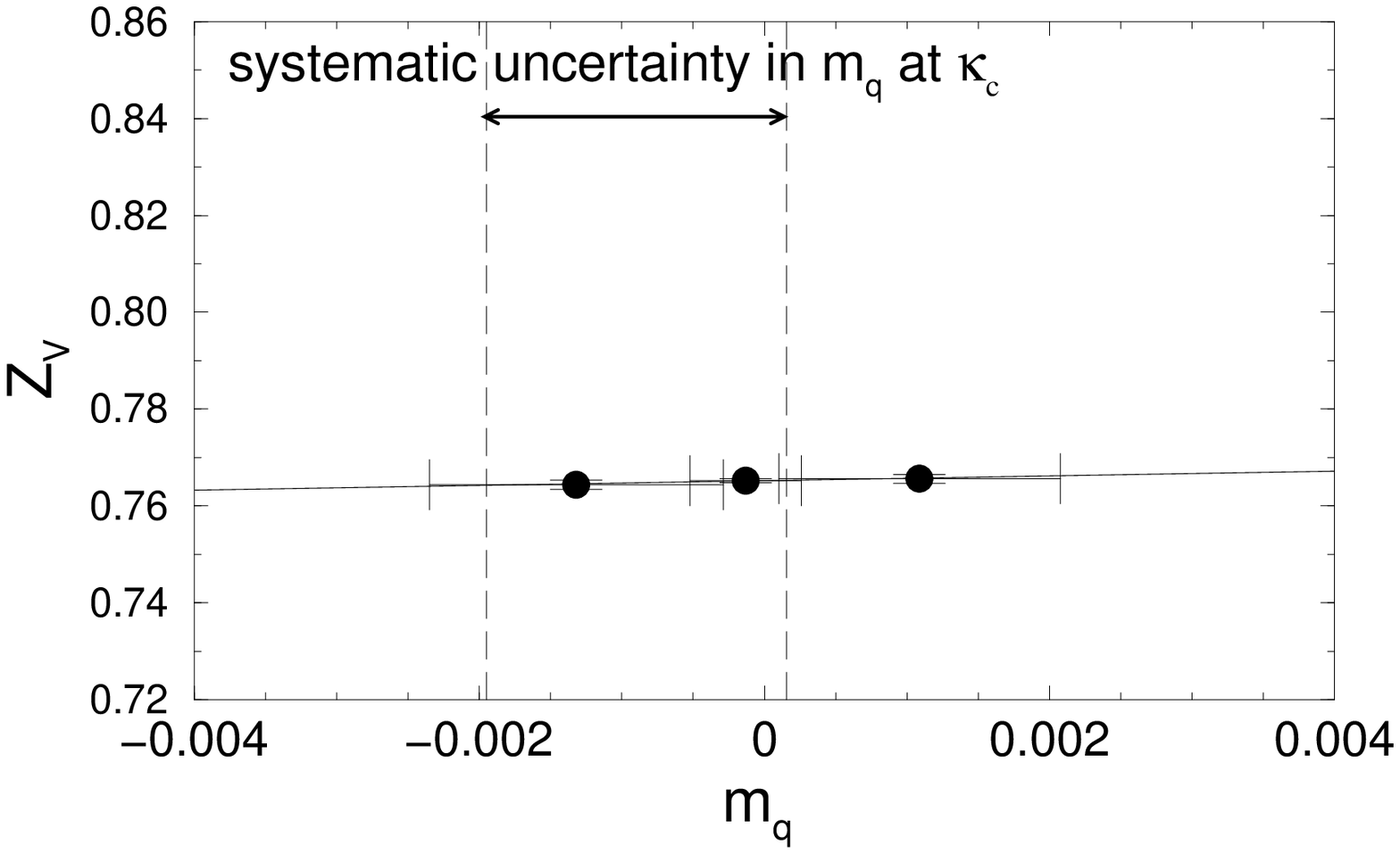}}
%\centerline{\vspace{6cm}}
\vspace{0.5cm}
\caption{$Z_V$ versus $m_q$ at $\beta=2.4$ on an $8^3\times 16$ lattice.}
\label{fig:m-Zv-2.4-8x16}
%\vspace{-0.7cm}
\end{figure}

\begin{figure}[htbp]
%\centerline{\epsfysize=10.0cm \epsfbox{eps/m-ZA-b=2.4.eps}}
\centerline{\epsfysize=10.0cm \epsfbox{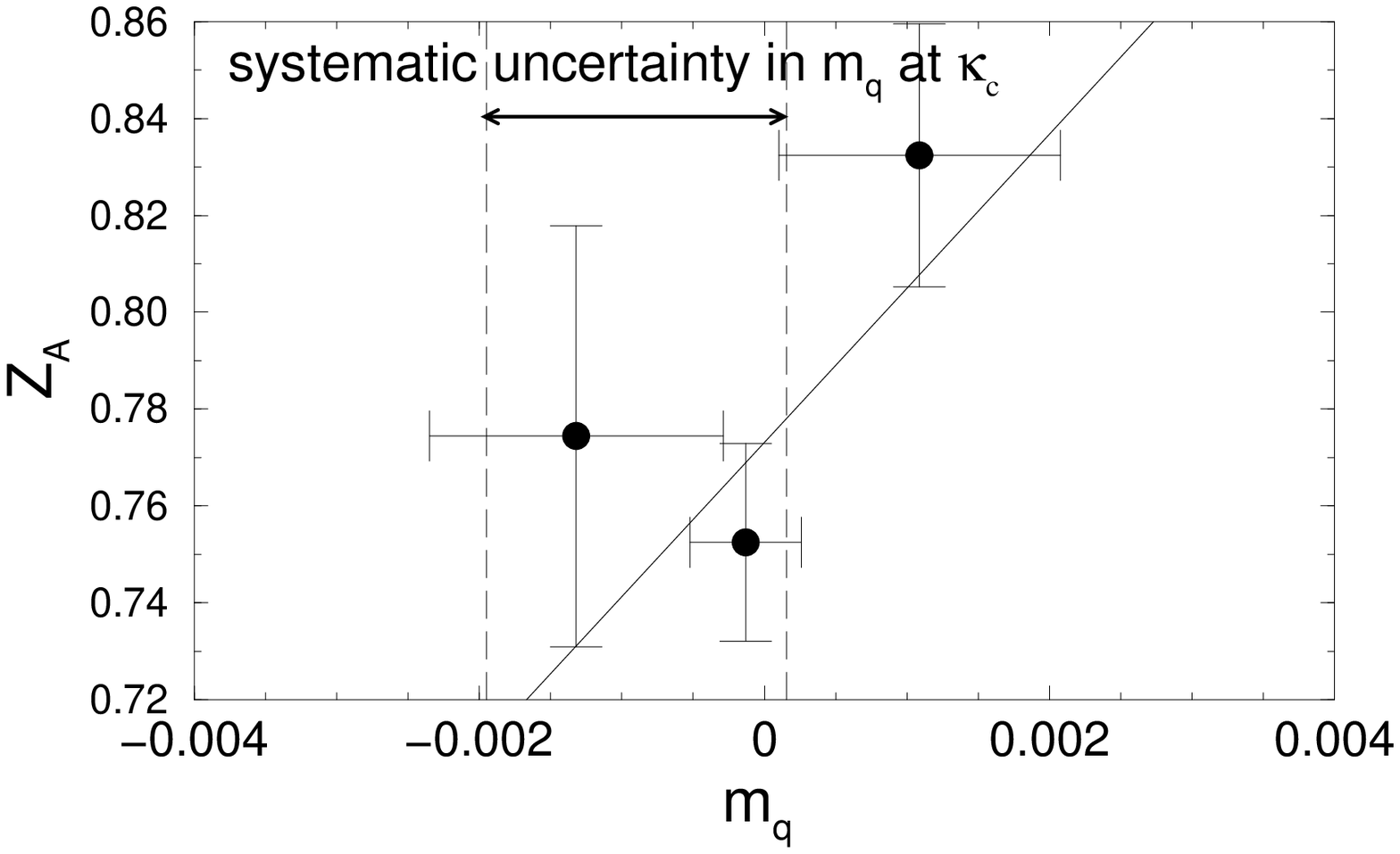}}
%\centerline{\vspace{6cm}}
\vspace{0.5cm}
\caption{$Z_A$ versus $m_q$ at $\beta=2.4$ on an $8^3\times 16$ lattice.}
\label{fig:m-ZA-2.4-8x16}
%\vspace{-0.7cm}
\end{figure}

%%%% Error of $Z$ from error of $m_q$ %%%% 
The uncertainty of $m_q$ is translated to
uncertainties of $Z$-factors.
To do this, we carry out two additional simulations at $\kappa$'s
slightly above and below $\kappa_c$. 
Figures~\ref{fig:m-Zv-2.4-8x16} and ~\ref{fig:m-ZA-2.4-8x16}
show how $Z$-factors depend on $m_q$;  
$Z_V$ is very stable against variation of $m_q$, while 
$Z_A$ shows a more prominent dependence.
We fit the $m_q$ dependence of $Z$-factors by a linear function, 
and uncertainties of $Z$-factors are estimated by the difference 
of the central value and the maximum/minimum value for 
the range of error of $m_q$. 
The uncertainties are given in Table~\ref{tab:ZVZA-error-estimate}
under the column $\delta Z^{m_q}$.

\begin{figure}[htbp]
%\centerline{\epsfysize=10.0cm \epsfbox{eps/f1-fv-Zv-2.4-8x16.eps}}
\centerline{\epsfysize=10.0cm \epsfbox{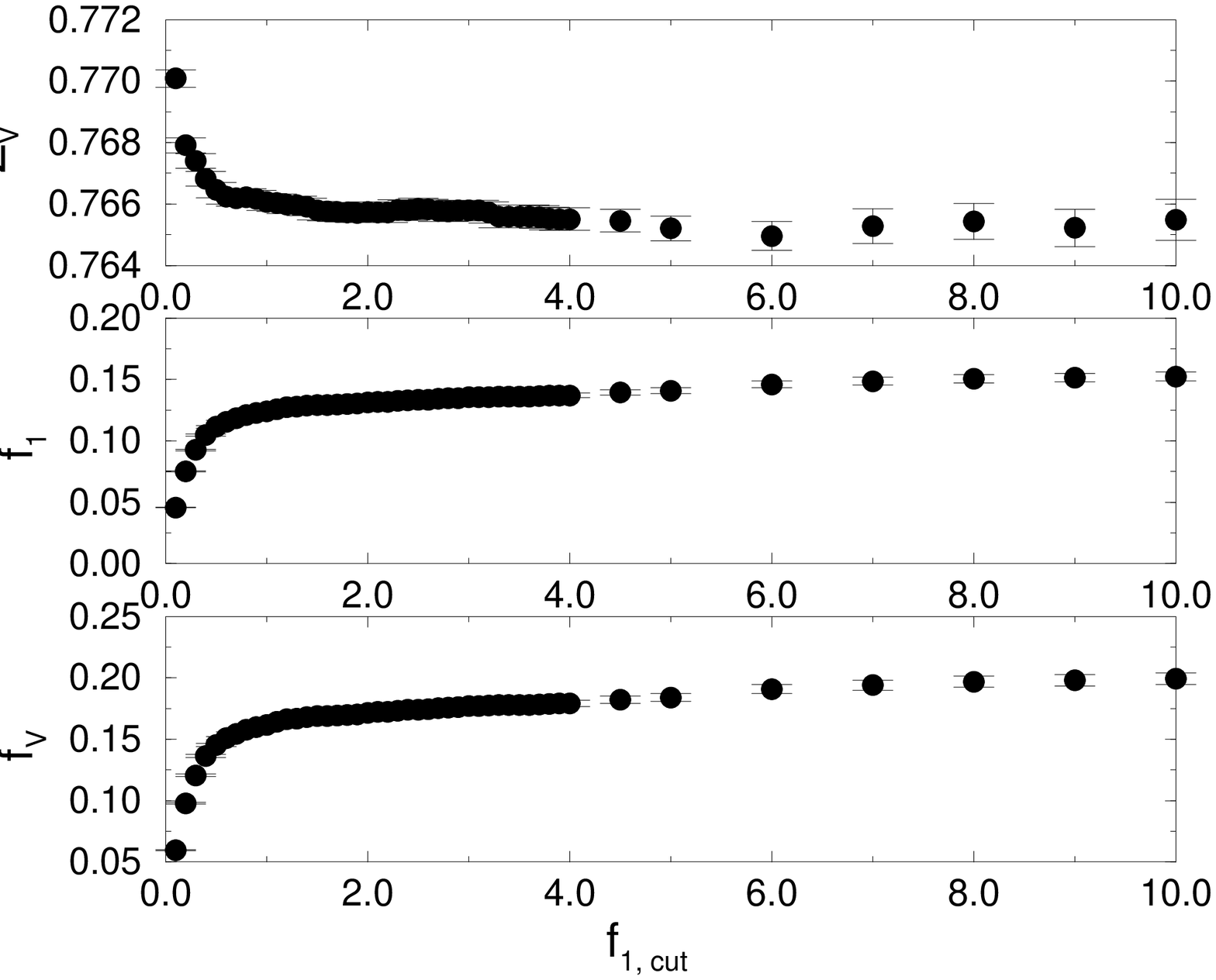}}
%\centerline{\vspace{6cm}}
\vspace{0.5cm}
\caption{Ensemble averages of $f_1$, $f_V$ and $Z_V$ versus 
$f_{1,{\rm cut}}$ at $\beta=2.4$ on an $8^3\times 16$ lattice.}
\label{fig:f1-fv-ZV.vs.fcut-2.4-8x16}
%\vspace{-0.7cm}
\end{figure}

\begin{figure}[htbp]
%\centerline{\epsfysize=10.0cm \epsfbox{eps/f1-fAA-ZA-2.4-8x16-rev.eps}}
\centerline{\epsfysize=10.0cm \epsfbox{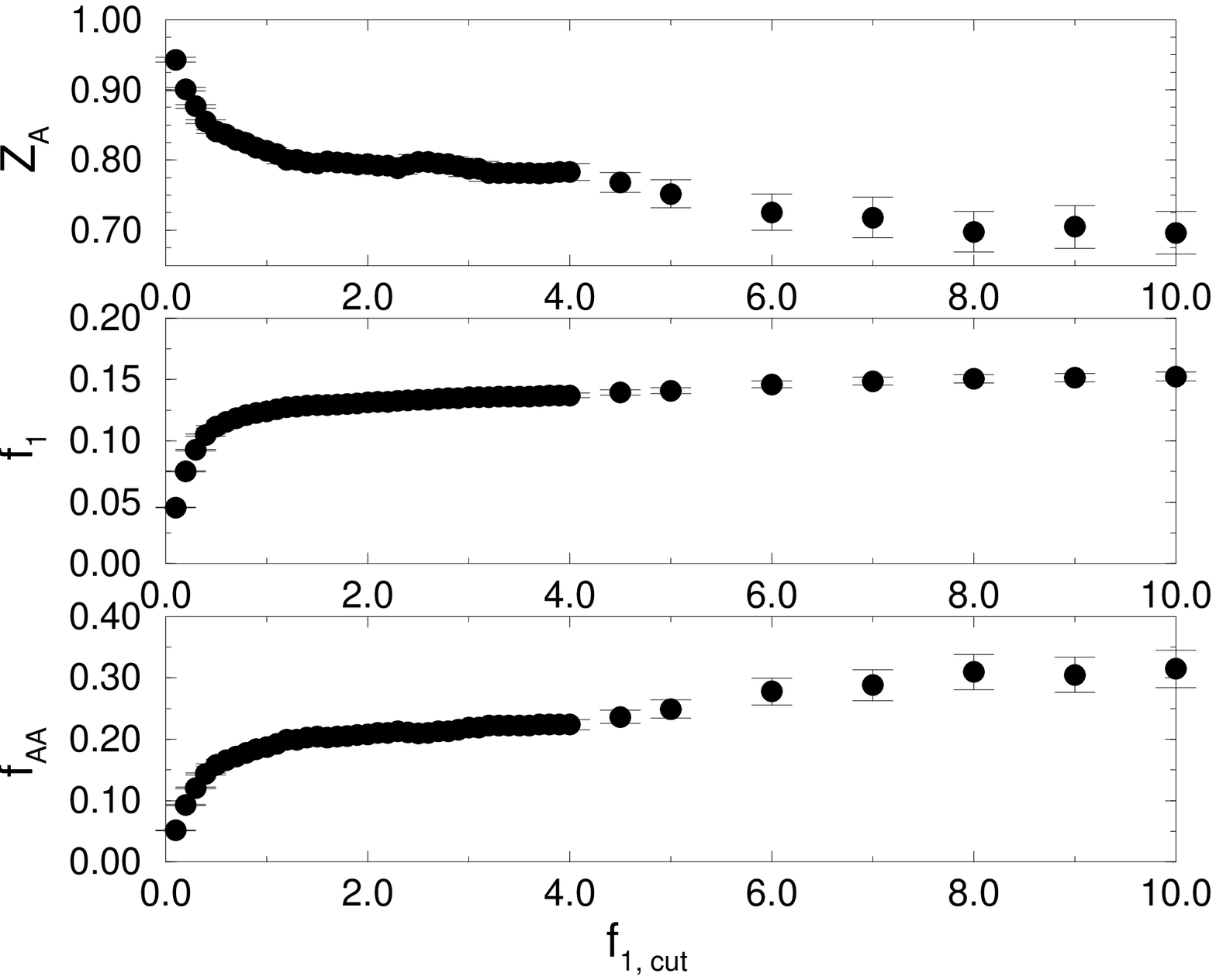}}
%\centerline{\vspace{6cm}}
\vspace{0.5cm}
\caption{Ensemble averages of $f_1$, $f_{AA}$ and $Z_A$ versus 
$f_{1,{\rm cut}}$ at $\beta=2.4$ on an $8^3\times 16$ lattice.}
\label{fig:f1-fAA-ZA.vs.fcut-2.4-8x16}
%\vspace{-0.7cm}
\end{figure}

%%%% $Z_V$ and $Z_A$ %%%%
We also estimate uncertainties in the statistical averaging 
of $Z$-factors themselves by varying the cutoff of $f_1$.
Figure~\ref{fig:f1-fv-ZV.vs.fcut-2.4-8x16} shows 
the ensemble average of $f_1$ and $f_V$ together with $Z_V$ 
as a function of the cutoff $f_{1,{\rm cut}}$.
Though both $f_1$ and $f_V$ increase as $f_{1,{\rm cut}}$, 
their ratio $Z_V$ is very stable around $f_{1,{\rm cut}}=5.0$ 
reflecting the fact that $f_1$ and $f_V$ are highly and positively
correlated.
We determine the central value from $f_{1,{\rm cut}}=5.0$ and
estimate errors by varying $f_{1,{\rm cut}}=2.0$ to 10.0.
Figure~\ref{fig:f1-fAA-ZA.vs.fcut-2.4-8x16} shows a similar 
test for $Z_A$.
The cutoff dependence is more conspicuous than for $Z_V$.
Uncertainties thus estimated are listed 
in Table~\ref{tab:ZVZA-error-estimate} under the column 
$\delta Z^{except.}$.
Note that we discard 47 (221) configurations at $\beta=2.4$ (2.2).
% 28 (144) for cut=10.0 at b=2.4 (2.2).

%%%% Total systematic error of $Z$ %%%%
The two uncertainties $\delta Z^{m_q}$ and $\delta Z^{except.}$
are simply added to estimate the total uncertainty
$\delta Z$ for $L/a=8$ lattices.
We then propagate them to uncertainties of $Z$-factors at
fixed physical volume, listed in Table~\ref{tab:ZVZA-error-estimate}.
Uncertainties at $L=0.8$~fm are smaller than those on $L/a=8$
lattices, because the physical size is located between $L/a=8$ and
$L/a=4$ lattices. 
Uncertainties of $Z$-factors at $L=\infty$, enlarged by 
extrapolations, are larger than the statistical error
$\delta Z^{stat,L=\infty}$.

\subsection{$Z$-factors at simulation points for meson decay constants}

%%%% What we want to determine %%%%
For discussions of scaling properties of meson decay constants, 
we need $Z$-factors at $\beta$ values where raw data of 
$f_\rho$ and $f_\pi$ are taken. 
We evaluate them using the Pad\'e fits obtained in sec.~\ref{subsec:resZ}
together with the estimates of uncertainties from exceptional configurations.

%%%% Error in Pade fits %%%%
For the latter purpose, we repeat Pad\'e fits varying 
the $Z$-factor at $\beta=2.4$ within the range of its uncertainty, calculate 
the $Z$-factor at a target $\beta$ value, and take the difference 
between this value and the central value as an estimate of the systematic error
from exceptional configurations at $\beta=2.4$. 
The systematic error from those at $\beta=2.2$ is estimated similarly
and added linearly to obtain the total systematic error.

%%%% Results %%%%
Our final results for $Z$-factors at the simulation points for meson 
decay constants are listed in Table.~\ref{tab:ZatDecayC}.
The systematic errors in $Z_V^{SF,L=\infty}$ and $Z_A^{SF,L=\infty}$ are 
comparable to the statistical ones and within at most 2$\sigma$.
The ratio of systematic errors to statistical ones in the table
is slightly smaller than the ratio at simulation points for
$Z$-factors, $\beta=2.4$ and 2.2, because the Pad\'e fits
are stabilized with data at $\beta \ge$ 2.6 where no exceptional
configurations appear.
In the table, we reproduce the one-loop tadpole improved values 
$Z_{V,A}^{TP}$ together with $Z_V^{\rm NPC}$ for comparison.

\section{Scaling properties of meson decay constants}\label{sec:decay}
%%%% $f_\rho$ %%%%
We investigate the scaling property of $f_\rho$, combining the unrenormalized
values of $f_\rho$~\cite{ref:CPPACS-NF2} and four different choices of $Z_V$;   
1) one-loop tadpole improved value, $Z_V^{TP}$, 
2) values determined by the SF method for $L=0.8$~fm,
$Z_V^{SF,L=0.8\rm{ fm}}$, 
3) those at $L=\infty$, $Z_V^{SF,L=\infty}$, and 
4) $Z_V^{\rm NPC}$ in eq.~(\ref{eq:ZNPC}) defined by the conserved current.
Unrenormalized $f_\rho$ was determined in Ref.~\cite{ref:CPPACS-NF2} 
for unimproved current. Hence, renormalized $f_\rho$ has $O(ag^2)$ error. 

Numerical values of the renormalized $f_\rho$ are listed 
in Table~\ref{tab:frhofpi}. 
Statistical errors in the unrenormalized $f_\rho$ and 
in $Z_V^{SF,L=0.8\rm{ fm}}$ or $Z_V^{SF,L=\infty}$ are added 
in quadrature, while systematic uncertainties for the latter are 
simply propagated to that of renormalized $f_\rho$.
Fortunately, systematic uncertainties from exceptional configurations
are sufficiently small. 
They are at most 0.4$\sigma$ of statistical errors and do not
affect the conclusions below.
Therefore we ignore systematic uncertainties in the following 
discussions.

\begin{figure}[htbp]
%\centerline{\epsfysize=10.0cm \epsfbox{eps/frho-Oa.eps}}
\centerline{\epsfysize=10.0cm \epsfbox{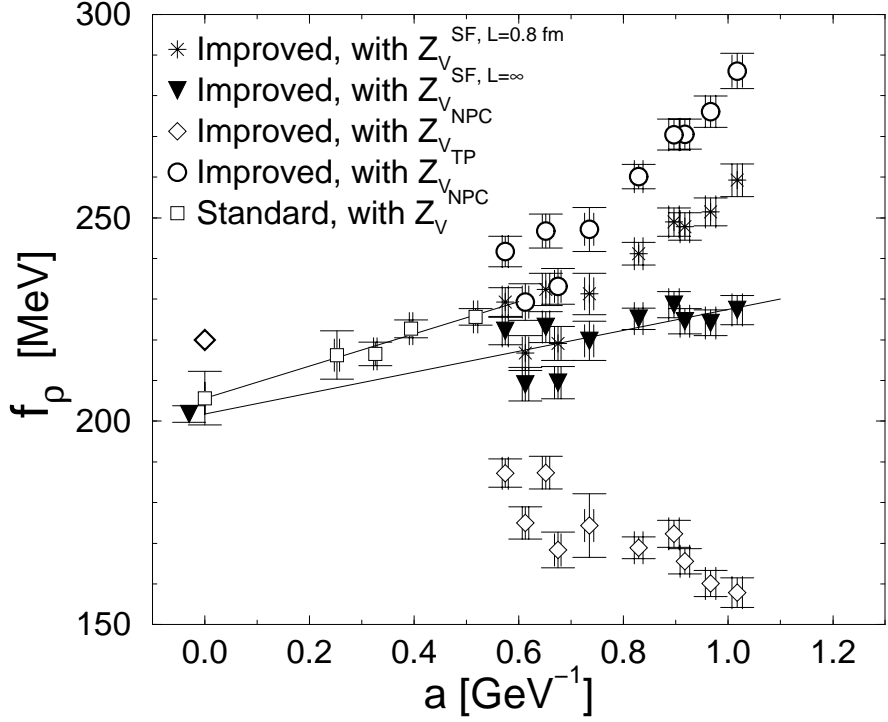}}
%\vspace{-1.1cm}
\caption{$f_\rho$ versus $a$ with various choices of $Z_V$.
Values for our improved action with $Z_V^{TP}$ and $Z_V^{\rm NPC}$
are taken from Ref.~\protect\cite{ref:CPPACS-NF2}.
Values for the standard action~\protect\cite{ref:CPPACS-quench} 
are determined with $Z$-factors from the conserved current.
Diamond at $a=0$ represents the experimental value.}
\label{fig:frho}
%\vspace{-0.7cm}
\end{figure}

Figure~\ref{fig:frho} presents the results for $f_\rho$ as a function of $a$. 
Open circles are the previous result obtained with tadpole improved 
one-loop $Z$-factor, which exhibits a sizable scaling violation. 
If instead one uses the non-perturbative $Z_V$ for $L=\infty$, 
we observe a much better scaling behavior (filled triangles). The $Z$-factor 
evaluated for $L=0.8$~fm lies in between the two results (stars). 

We find this result to be very encouraging; it shows that an apparent large 
scaling violation seen with the use of $Z_V^{TP}$ is largely due to 
neglect of higher order contributions in the $Z$-factor, and that 
inherent $O(a)$ errors in the vector decay constant are small even 
up to a large lattice spacing of $a^{-1}\sim 1$~GeV.

%Scaling is best improved when $Z_V$ is fixed for $L=\infty$.  
%This property is likely related to the fact that $O(ag^4/L)$ errors
%in $Z_V$ are removed by the limiting procedure $L\to\infty$. 
%(We note that $O(a\Lambda)$ error still remains in $Z_V^{SF,L=\infty}$
%and also in unrenormalized $f_\rho$.)

We extrapolate $f_\rho$ with $Z_V^{SF,L=\infty}$ to 
the continuum limit linearly in $a$. 
% since our action has $O(ag^4)$
%error and unrenormalized $f_\rho$ is determined for unimproved operator.
The value in the continuum limit $f_\rho=201.7(2.0)$~MeV turns out to be 
consistent with that $f_\rho=205.7(6.6)$~MeV from a high precision simulation with the
standard action~\cite{ref:CPPACS-quench}.

We note that a relatively large scaling violation is observed for
$f_\rho$ with $Z_V^{\rm NPC}$ (open diamonds).
Since the $f_\rho$ with $Z_V^{\rm NPC}$ is nothing but decay constant
determined from the conserved current in eq.~(\ref{eq:VC}),
we suspect that $O(a)$ large scaling violation exists in the 
conserved current.

%%%% $f_\pi$ %%%%
The scaling property of $f_\pi$ is investigated in a similar manner.
We note that unrenormalized $f_\pi$ in Ref.~\cite{ref:CPPACS-NF2} 
was determined for the improved current using a tadpole improved 
one-loop value of $c_A$ with $\overline{\mbox{MS}}$ coupling, 
whereas $Z$-factors in this work employ $c_A$ with bare coupling.
Difference in renormalized $f_\pi$ arising from the difference of
$c_A$ remains $O(ag^4)$.

\begin{figure}[htbp]
%\centerline{\epsfysize=10.0cm \epsfbox{eps/fpi-Oa.eps}}
\centerline{\epsfysize=10.0cm \epsfbox{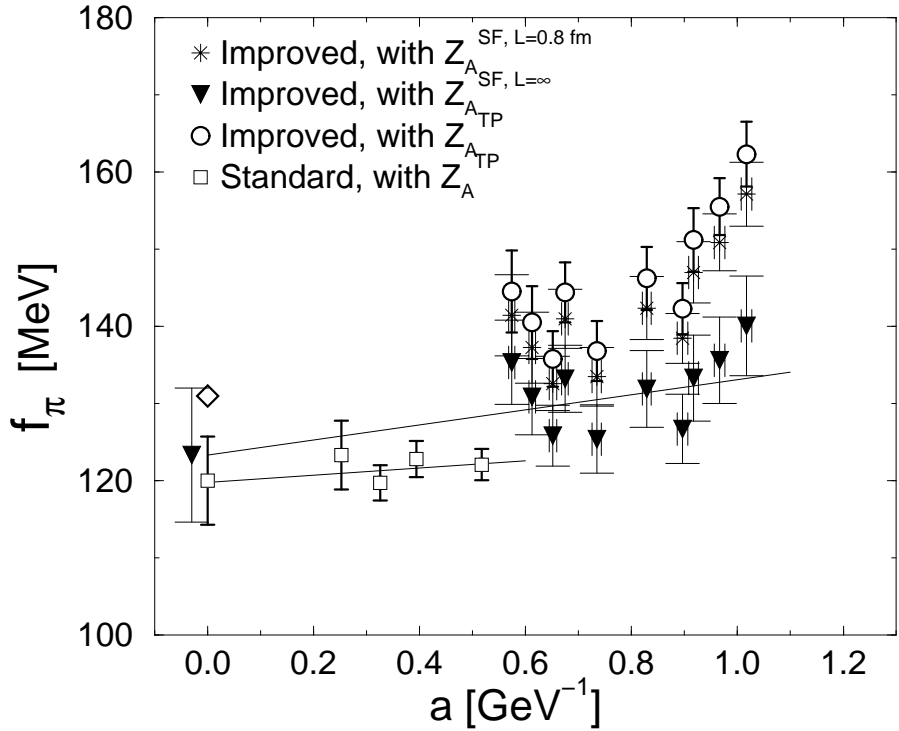}}
%\vspace{-1.1cm}
\caption{$f_\pi$ versus $a$ with various choices of $Z_A$.
Symbols are the same as in Fig.~\protect\ref{fig:frho}, 
though $f_\pi$ from the standard action is determined with
$Z_A^{TP}$.}
\label{fig:fpi}
%\vspace{-0.7cm}
\end{figure}

Figure~\ref{fig:fpi} shows $f_\pi$ versus $a$. In this figure, we have not 
included the systematic uncertainty since, being smaller than statistical 
one, it does not alter the conclusions below.
Numerical values are listed in Table~\ref{tab:frhofpi}.  
The conclusions on scaling behavior are similar to those of 
$f_\rho$.  Employing $Z_A^{SF,L=\infty}$, which includes terms to all 
orders in $g^2$ and does not have $O(ag^4/L)$ error, 
scaling behavior turns out to be better than 
previous results with tadpole improved one-loop $Z$-factor indicated. 
Making a linear continuum extrapolation, we obtain a value $f_\pi=123.3(8.7)$~MeV consistent 
with that $f_\pi=120.0(5.7)$~MeV calculated with the standard action at significantly weaker 
couplings (open squares). 
%In the figure, we do not include systematic uncertainty of $f_\pi$. 
%As far as we ignore the systematic uncertainty, the scaling property of
%$f_\pi$ is similar to the case of $f_\rho$. 
%Namely, scaling violation observed for $f_\pi$ with $Z_A^{TP}$ is
%reduced by the use of $Z_A^{SF,L=\infty}$, and the value in the
%continuum limit of $f_\pi$ with $Z_A^{SF,L=\infty}$ is consistent
%with that from the standard action.
%We note, however, that $f_\pi$ with $Z_A^{SF,L=\infty}$ has
%large statistical error and systematic one of order 0.5$\sigma$
%at $a=$ 1 GeV$^{-1}$.
%We conclude that scaling property is improved with
%$Z_A^{SF,L=\infty}$, because $f_\pi$ with $Z_A^{SF,L=\infty}$
%around $a=$ 1 GeV$^{-1}$ is smaller than that with $Z_A^{TP}$
%beyond the sum of statistical and systematic errors, and is
%closer to the value in the continuum limit.

\section{Conclusions}\label{sec:conclusion} 

The Schr\"odinger functional method has been applied to 
calculations of vector and axial-vector renormalization constants
for the combination of a RG-improved gauge action and 
a tadpole-improved clover quark action.
With the $Z$-factors determined non-perturbatively, 
an apparent large scaling violation in the range of
lattice spacing $a^{-1}=1-2$~GeV in the 
meson decay constants previously observed with one-loop perturbative 
$Z$-factors is significantly reduced. 
We conclude that the improvement attempted with the 
gluon and quark actions employed in the present work 
is effective for the meson decay constants as well. 
%the large scaling violation observed
%in Ref.~\cite{ref:CPPACS-NF2} originates mainly from scaling
%violations of perturbative $Z$-factors.

We find that scaling of decay constants is best when one uses 
$Z$-factors normalized at infinite volume. 
This suggests that removing $O(ag^4/L)$ error in $Z$-factors by 
the limiting procedure $L\to\infty$ is important for actions
with $O(ag^4)$ error.

The non-perturbative $Z$-factors have enabled us to determine
values in the continuum limit of decay constants.
We may expect that other hadronic matrix elements are also
reliably extracted from lattice spacings much coarser than
$a^{-1}\approx$ 2 GeV for our action combination if one uses 
$Z$-factors determined by the Schr\"odinger functional method.   

%Finally we recall that large scaling violation has been 
%observed for the meson decay constants also in two-flavor full 
%QCD with the same action combinations as used in this article for 
%quenched QCD.
Finally we recall that a large scaling violation for the meson decay constants
has been observed also in two-flavor full QCD with the same action combination
of quenched QCD considered in this article.
Non-perturbative determination of $Z$-factors will 
therefore be interesting to pursue for this case. 
Exceptional configurations are expected to be absent in full QCD. 
Therefore the Schr\"odinger functional calculation would be more
straightforward. Work in this direction is in progress.

\acknowledgements

This work is supported in part by Grants-in-Aid of the Ministry of Education 
(Nos.~
12304011, % iwasaki
%12640253, % saoki
%12740133, % ishizuka
13135204, % iwasaki
13640260, % kanaya
14046202, % saoki
14740173, % kaneko
15204015, % ukawa
15540251, % saoki
15540279  % okawa
16028201  % saoki
). 
%SE and M. Okamoto are JSPS Research Fellows. 
\comment{VL is supported by the Research for Future Program of JSPS
(No. JSPS-RFTF 97P01102).}

\newpage
%%%%%%%%%%%%%% tables %%%%%%%%%%%%%%%%

\begin{table}
\caption{Simulation parameters and $Z$-factors at simulation points.
For $8^3\times 16$ lattices at $\beta=2.2$ and 2.4 where we encounter
exceptional configurations, we quote only central values and
statistical errors. No calculations of $Z_A$ were made 
for $16^3\times 32$ lattice at $\beta=2.8$ and $12^3\times 24$ lattice
at $\beta=8.0$.}
\label{tab:kcmqZVZAatSP}
\begin{tabular}{lrrllrlr}
\ $\beta$ & $L^3\times T$ & \multicolumn{1}{c}{$\kappa_c$} & \multicolumn{1}{c}{$m_q$} &
	\multicolumn{1}{c}{$Z_V$} & \#conf for $Z_V$ &
	\multicolumn{1}{c}{$Z_A$} & \#conf for $Z_A$ \\\hline
2.2   & $ 4^3 \times\hspace{1.6mm}8$ & 0.139281(56) & -0.000529(495) & 0.7499(11) & 2000 & 0.7667(52) & 2000 \\
      & $ 8^3 \times 16$             & 0.140570(15) &\ 0.001493(399) & 0.7067(12) &20000 & 0.7350(610)&20000 \\
2.4   & $ 4^3 \times\hspace{1.6mm}8$ & 0.136933(21) & -0.000764(337) & 0.7995(10) & 1000 & 0.8150(24) & 2000 \\
      & $ 8^3 \times 16$             & 0.137481(04) & -0.000130(395) & 0.7652(04) &10000 & 0.7525(204)&10000 \\
2.6   & $ 4^3 \times\hspace{1.6mm}8$ & 0.135558(08) & -0.000276(341) & 0.8265(05) & 2000 & 0.8474(16) & 2000 \\
      & $ 8^3 \times 16$             & 0.135701(10) & -0.000464(154) & 0.8056(05) &  500 & 0.8208(18) & 2000 \\
2.8   & $ 4^3 \times\hspace{1.6mm}8$ & 0.134532(10) &\ 0.000002(396) & 0.8482(06) & 1000 & 0.8667(18) & 1000 \\
      & $ 8^3 \times 16$             & 0.134515(07) & -0.000105(139) & 0.8312(04) &  500 & 0.8482(13) & 1000 \\
      & $12^3 \times 24$             & 0.134554(08) &\ 0.000525(111) & 0.8268(03) &  500 & 0.8520(17) & 1000 \\
      & $16^3 \times 32$             & 0.134587(09) &\ 0.000663(139) & 0.8231(06) &  500 &            &      \\
3.125 & $12^3 \times 24$             & 0.133209(01) & -0.000039(091) & 0.8540(02) &  500 & 0.8689(09) &  500 \\
      & $16^3 \times 32$             & 0.133219(05) & -0.000092(060) & 0.8527(01) &  500 & 0.8672(08) &  500 \\
%     & $16^3 \times 48$             &              &                &            &      & 0.8720(44) &  500 \\
4.0   & $ 8^3 \times 16$             & 0.131094(04) & -0.000302(086) & 0.8972(02) &  500 & 0.9068(06) &  500 \\
%     & $ 8^3 \times 24$             &              &                &            &      & 0.9012(14) &  500 \\
      & $16^3 \times 32$             & 0.131083(01) &\ 0.000160(044) & 0.8938(01) &  500 & 0.9060(04) &  300 \\
%     & $16^3 \times 48$             &              &                &            &      & 0.9091(09) &  500 \\
6.0   & $ 8^3 \times 16$             & 0.128898(03) & -0.000191(061) & 0.9366(01) &  300 & 0.9425(03) &  300 \\
%     & $ 8^3 \times 24$             &              &                &            &      & 0.9424(07) &  300 \\
      & $16^3 \times 32$             & 0.128891(01) & -0.000060(023) & 0.9337(01) &  300 & 0.9398(02) &  300 \\
%     & $16^3 \times 48$             &              &                &            &      & 0.9397(04) &  300 \\
8.0   & $ 8^3 \times 16$             & 0.127869(02) &\ 0.000138(041) & 0.9545(01) &  300 & 0.9600(02) &  300 \\
%     & $ 8^3 \times 24$             &              &                &            &      & 0.9598(04) &  300 \\
      & $12^3 \times 24$             & 0.127870(01) &\ 0.000576(331) & 0.9523(01) &  300 &            &      \\
      & $16^3 \times 32$             & 0.127870(01) &\ 0.000021(017) & 0.95151(3) &  300 & 0.9565(01) &  200 \\
%     & $16^3 \times 48$	     &              &                &            &      & 0.9558(02) &  300 \\
\end{tabular}
\end{table}

\begin{table}
\caption{Z-factors at fixed physical volumes.}
\label{tab:ZVZA-phys}
\begin{tabular}{lllll}
\ $\beta$ & $Z_V^{SF,L=0.8\rm{ fm}}$  & $Z_V^{SF,L=\infty}$ &
	    $Z_A^{SF,L=0.8\rm{ fm}}$  & $Z_A^{SF,L=\infty}$ \\\hline
2.2   & 0.7495(15) & 0.6635(26) & 0.7664(52) & 0.7033(1221)\\
2.4   & 0.7783(05) & 0.7309(13) & 0.7763(127)& 0.6900(409) \\
2.6   & 0.8056(05) & 0.7847(11) & 0.8208(18) & 0.7942(39)  \\
2.8   & 0.8279(02) & 0.8180(12) & 0.8511(13) & 0.8386(22)  \\
3.125 & 0.8526(01) & 0.8488(07) & 0.8671(08) & 0.8621(42)  \\
4.0   & 0.8917(02) & 0.8904(03) & 0.9055(08) & 0.9052(10)  \\
6.0   & 0.9309(02) & 0.9308(02) & 0.9372(05) & 0.9371(05)  \\
8.0   & 0.9485(01) & 0.9485(01) & 0.9530(03) & 0.9530(03)  \\
\end{tabular}
\end{table}

\begin{table}
\caption{Breakdown of systematic uncertainties in $Z$-factors.
See text for details.}
\label{tab:ZVZA-error-estimate}
\begin{tabular}{lllllll}
  & $\delta Z^{m_q}$ & $\delta Z^{except.}$ & $\delta Z$ & $\delta
Z^{SF,L=0.8\rm{ fm}}$ & $\delta Z^{SF,L=\infty}$ & $\delta Z^{stat,L=\infty}$\\\hline
$Z_V (\beta=2.2)$  & +0.04\%$,-$0.02\%  & +0.31\%,$-$0.06\% 
                   & +0.35\%,$-$0.08\% 
                   & +0.0013\%,$-$0.0058\% & +0.75\%,$-$0.17\% & 0.39\%\\
$Z_V (\beta=2.4)$  & +0.01\%,$-$0.12\%  & +0.08\%,$-$0.03\% & +0.09\%,$-$0.15\% 
                   & +0.044\%,$-$0.10\%  & +0.19\%,$-$0.31\% & 0.18\%\\
$Z_A (\beta=2.2)$  & +3.2\%,$-$1.5\% & +18.1\%,$-$2.1\%  & +21.3\%,$-$3.6\% 
                   & +0.17\%,$-$0.25\%   & +44.5\%,$-$8.9\% & 17\%\\
$Z_A (\beta=2.4)$  & +0.6\%,$-$8.0\% & \ +6.1\%,$-$7.3\% & +6.7\%,$-$15.3\% 
                   & +4.0\%,$-$9.2\% \ & +14.6\%,$-$33.4\% & 5.9\% \\
\end{tabular}
\end{table}

\begin{table}
\caption{$Z$-factors at simulation points for meson decay constants.}
\label{tab:ZatDecayC}
\begin{tabular}{llllllll}
$\beta$ & $Z_V^{TP}$ & $Z_V^{SF,L=0.8\rm{ fm}}$ &  $Z_V^{SF,L=\infty}$ & $Z_V^{\rm NPC}$ 
        & $Z_A^{TP}$ & $Z_A^{SF,L=0.8\rm{ fm}}$ &  $Z_A^{SF,L=\infty}$ \\\hline
2.187 & 0.81657 & 0.7400(08)$^{+04}_{-10}$ & 0.6488(24)$^{+48}_{-27}$ & 0.4536(044) & 0.83204 & 0.7712(38)$^{+27}_{-54}$ & 0.7180(146)$^{+125}_{-236}$\\
2.214 & 0.81923 & 0.7462(07)$^{+04}_{-09}$ & 0.6651(20)$^{+40}_{-23}$ & 0.4784(038) & 0.83449 & 0.7762(34)$^{+24}_{-49}$ & 0.7276(127)$^{+109}_{-203}$\\
2.247 & 0.82237 & 0.7532(06)$^{+03}_{-08}$ & 0.6825(16)$^{+32}_{-19}$ & 0.5056(030) & 0.83737 & 0.7819(30)$^{+21}_{-43}$ & 0.7383(108)$^{+093}_{-171}$\\
2.281 & 0.82548 & 0.7601(06)$^{+03}_{-07}$ & 0.6981(13)$^{+26}_{-16}$ & 0.5293(040) & 0.84022 & 0.7875(27)$^{+19}_{-38}$ & 0.7482(092)$^{+080}_{-143}$\\
2.334 & 0.83009 & 0.7698(04)$^{+02}_{-06}$ & 0.7187(10)$^{+20}_{-13}$ & 0.5421(030) & 0.84446 & 0.7956(22)$^{+16}_{-31}$ & 0.7619(073)$^{+063}_{-111}$\\
2.416 & 0.83673 & 0.7832(03)$^{+02}_{-04}$ & 0.7441(07)$^{+13}_{-09}$ & 0.5925(177) & 0.85057 & 0.8068(17)$^{+12}_{-23}$ & 0.7797(052)$^{+045}_{-076}$\\
2.456 & 0.83978 & 0.7891(03)$^{+01}_{-03}$ & 0.7544(06)$^{+10}_{-08}$ & 0.6090(043) & 0.85336 & 0.8118(15)$^{+10}_{-20}$ & 0.7873(045)$^{+038}_{-064}$\\
2.487 & 0.84205 & 0.7934(03)$^{+01}_{-03}$ & 0.7617(06)$^{+09}_{-07}$ & 0.6420(038) & 0.85546 & 0.8155(14)$^{+09}_{-18}$ & 0.7927(040)$^{+034}_{-056}$\\
2.528 & 0.84496 & 0.7988(02)$^{+01}_{-03}$ & 0.7705(05)$^{+07}_{-06}$ & 0.6460(038) & 0.85813 & 0.8201(12)$^{+08}_{-15}$ & 0.7994(034)$^{+029}_{-047}$\\
2.575 & 0.84816 & 0.8046(02)$^{+01}_{-02}$ & 0.7796(05)$^{+06}_{-05}$ & 0.6604(034) & 0.86107 & 0.8250(11)$^{+07}_{-13}$ & 0.8064(029)$^{+024}_{-039}$\\
\end{tabular}
\end{table}

\begin{table}
\caption{$f_\rho$ and $f_\pi$ in GeV with various choices of $Z$-factors.}
\label{tab:frhofpi}
\begin{tabular}{lllllllll}
$\beta$ & $a^{-1}[GeV]$ & $f_\rho(Z_V^{TP})$ & $f_\rho(Z_V^{SF,L=0.8 \rm{ fm}})$ & $f_\rho(Z_V^{SF,L=\infty})$ & $f_\rho(Z_V^{\rm NPC})$ & $f_\pi(Z_A^{TP})$ & $f_\pi(Z_A^{SF,L=0.8 \rm{ fm}})$ & $f_\pi(Z_A^{SF,L=\infty})$
\\\hline 
2.187 & 1.017(10) & 0.2861(44) & 0.2593(43)$^{+01}_{-04}$ & 0.2273(43)$^{+17}_{-09}$ & 0.1578(37) & 0.1623(42) & 0.1504(46)$^{+05}_{-11}$ & 0.1401(65)$^{+24}_{-46}$\\
2.214 & 0.966(10) & 0.2761(38) & 0.2515(37)$^{+01}_{-03}$ & 0.2242(38)$^{+13}_{-08}$ & 0.1601(32) & 0.1555(37) & 0.1446(41)$^{+04}_{-09}$ & 0.1356(56)$^{+20}_{-38}$\\
2.247 & 0.917(09) & 0.2706(37) & 0.2478(36)$^{+01}_{-03}$ & 0.2246(36)$^{+11}_{-06}$ & 0.1656(31) & 0.1512(41) & 0.1412(44)$^{+04}_{-08}$ & 0.1333(56)$^{+17}_{-31}$\\
2.281 & 0.896(10) & 0.2704(38) & 0.2490(37)$^{+01}_{-02}$ & 0.2287(36)$^{+09}_{-05}$ & 0.1723(33) & 0.1423(33) & 0.1334(36)$^{+03}_{-06}$ & 0.1267(45)$^{+14}_{-24}$\\
2.334 & 0.829(08) & 0.2601(30) & 0.2412(29)$^{+01}_{-02}$ & 0.2252(29)$^{+06}_{-04}$ & 0.1689(27) & 0.1462(41) & 0.1377(42)$^{+03}_{-05}$ & 0.1319(50)$^{+11}_{-19}$\\
2.416 & 0.734(09) & 0.2471(54) & 0.2313(51)$^{+01}_{-01}$ & 0.2197(50)$^{+04}_{-03}$ & 0.1744(78) & 0.1368(39) & 0.1298(40)$^{+02}_{-04}$ & 0.1254(44)$^{+07}_{-12}$\\
2.456 & 0.674(06) & 0.2332(44) & 0.2191(42)$^{+00}_{-01}$ & 0.2095(41)$^{+03}_{-02}$ & 0.1683(44) & 0.1444(39) & 0.1374(40)$^{+02}_{-03}$ & 0.1332(44)$^{+06}_{-11}$\\
2.487 & 0.652(07) & 0.2467(42) & 0.2324(40)$^{+00}_{-01}$ & 0.2232(40)$^{+03}_{-02}$ & 0.1874(40) & 0.1358(36) & 0.1295(37)$^{+01}_{-03}$ & 0.1258(40)$^{+05}_{-09}$\\
2.528 & 0.612(06) & 0.2293(45) & 0.2168(43)$^{+00}_{-01}$ & 0.2091(42)$^{+02}_{-02}$ & 0.1750(40) & 0.1405(47) & 0.1343(47)$^{+01}_{-02}$ & 0.1309(49)$^{+05}_{-08}$\\
2.575 & 0.574(06) & 0.2417(37) & 0.2293(36)$^{+00}_{-01}$ & 0.2222(35)$^{+02}_{-01}$ & 0.1872(34) & 0.1445(53) & 0.1384(53)$^{+01}_{-02}$ & 0.1353(55)$^{+04}_{-06}$\\
\end{tabular}
\end{table}

\newpage

%%%%%%%%%%%%%% figures %%%%%%%%%%%%%%%%

\end{document}